\documentclass[twocolumn]{autart}    

\usepackage{graphicx}          
\usepackage{times} 
\usepackage{amsmath} 
\usepackage{amssymb}  
\usepackage{subcaption}
\usepackage{setspace}

\usepackage{xcolor}

\begin{document}

\begin{frontmatter}

\title{On compression rate of quantum autoencoders: Control design, numerical and experimental realization\thanksref{footnoteinfo}} 

\thanks[footnoteinfo]{This work was supported by the National Natural Science Foundation of
China (No. 61833010 and No. 61773232), the Australian Research
Council's Discovery Projects funding scheme under Project DP190101566, U.S. office of Naval Research Global under Grant N62909-19-1-2129, and the National Key R$\&$D Program of China
(No. 2018YFA0306703 and No. 2017YFA0304304).}
\author[adfa]{Hailan Ma}\ead{hailanma0413@gmail.com},    
\author[ustc1,ustc2]{Chang-Jiang Huang}\ead{hcj00@mail.ustc.edu.cn},
\author[nju]{Chunlin Chen}\ead{clchen@nju.edu.cn},
\author[adfa,b]{Daoyi Dong}\ead{daoyidong@gmail.com},  
\author[griffith]{Yuanlong Wang}\ead{yuanlong.wang.qc@gmail.com},    
\author[Tsing]{Re-Bing Wu}\ead{rbwu@tsinghua.edu.cn},
\author[ustc1,ustc2,b]{Guo-Yong Xiang}\ead{gyxiang@ustc.edu.cn}

\thanks[b]{Corresponding author.}

\address[adfa]{School of Engineering and Information Technology, University of New South Wales, Canberra, ACT 2600, Australia}
\address[ustc1]{Key Laboratory of Quantum Information, University of Science and Technology of China, CAS, Hefei 230026, China}
\address[ustc2]{CAS Center for Excellence in Quantum Information and Quantum Physics}
\address[nju]{Department of Control and Systems Engineering, School of Management and Engineering, Nanjing University, Nanjing
210093, China}
\address[griffith]{Centre for Quantum Dynamics, Griffith University, Brisbane, Queensland 4111, Australia}
\address[Tsing]{Department of Automation, Tsinghua University, Beijing 100084, China}


\begin{abstract}                          
Quantum autoencoders which aim at compressing quantum information in a low-dimensional latent space lie in the heart of automatic data compression in the field of quantum information.
In this paper, we establish an upper bound of the compression rate for a given
quantum autoencoder and present a learning control approach for training the
autoencoder to achieve the maximal compression rate. The upper bound of the
compression rate is theoretically proven using eigen-decomposition and matrix differentiation, which is determined by the eigenvalues of the density matrix
representation of the input states.
Numerical results on 2-qubit and 3-qubit systems are presented to demonstrate how to train the quantum autoencoder to achieve the theoretically  maximal compression, and the training performance using different machine learning algorithms is compared. Experimental results of a quantum autoencoder using quantum optical systems are illustrated for compressing two 2-qubit states into two 1-qubit states.
\end{abstract}

\end{frontmatter}

\section{Introduction}
In the recent decades, quantum technologies have attracted considerable attention towards realizing a variety of promising goals \cite{dong2010quantum}, \cite{nielsen2010quantum} such as implementing practical quantum
computers, building quantum communication networks, manipulating quantum systems, and developing quantum sensors  \cite{ticozzi2009analysis}, \cite{amini2013feedback},
\cite{zhang2020dynamics}, \\
\cite{bao2021fundamental}, \cite{dong2022quantum}, \\ \cite{bao2022exponentially}. A limitation for these applications is the capacity of quantum
resources, such as quantum coherence, which can be challenging to generate,
manipulate and preserve effectively. As such, quantum data
compression has been investigated \cite{bartuuvskova2006optical}, \cite{datta2013one}, \cite{rozema2014quantum} to reduce
the requirements on quantum memory and quantum communication
channels \cite{steinbrecher2019quantum}, and decrease the size of quantum
gates \cite{lamata2018quantum}, \cite{ding2019experimental}. It may benefit
various applications including quantum simulation \cite{aspuru2005simulated}, quantum communication and
distributed computation in a quantum network \cite{steinbrecher2019quantum}, \\ \cite{lamata2018quantum}. In these applications, it is highly desirable
to find universal tools to reduce the overhead in terms of valuable quantum
resources.



In the field of classical data processing, autoencoders are powerful
approaches for dimension reduction \cite{baldi2012autoencoders}, and have wide applications in different control tasks. For example, an empirical model which encapsulates measured data into a few parameters helps achieve an enhanced estimate of physical parameters  \cite{gawthrop2005data}. Autoencoders have been introduced to learn latent representations of nonlinear state-space model for system identification problems \cite{masti2018learning}, \cite{masti2021learning} and have been applied to solve the model identification issues from video data \cite{beintema2021non}. Recently, autoencoders for quantum data have received much attention in the field of quantum information.
For example, Wan \emph{et al.} \cite{wan2017quantum} introduced a feedforward quantum
neural network for quantum data compression. Romero \emph{et al.} \cite{romero2017quantum} formulated a simple autoencoder framework using a programmable circuit and applied it to compress the ground states of the Hubbard model and molecular Hamiltonians. For dimension reduction of qudits, Pepper \emph{et al.} \cite{pepper2019experimental} utilized the occupation probability of ``junk" mode as the cost function and experimentally realized a quantum autoencoder with low-level errors. Huang \emph{et al.} \cite{huang2020experimental} experimentally realized a 2-qubit compression scheme, where information originated from a path qubit and
a polarization qubit was encoded in the polarization qubit. Steinbrecher \emph{et al.} \cite{steinbrecher2019quantum} designed a quantum optical neural network and
implemented simulations for quantum optical state compression. Lamata \emph{et al.} \cite{lamata2018quantum} analyzed the connection between a quantum autoencoder
and an approximate quantum adder, and proposed an approach that employed a
restricted number of gates to achieve a quantum autoencoder with high
fidelity. Its experimental implementation in the Rigetti cloud quantum computer achieved fidelities in good agreement with
theoretical predictions \cite{ding2019experimental}. Recently, quantum autoencoders have been investigated for different applications. For example, a novel quantum
autoencoder has been developed to denoise Greenberger-Horne-Zeilinger (GHZ) states subject to spin-flip errors and random unitary noise \cite{bondarenko2020quantum}. A detection-based error mitigation using quantum autoencoders has been successfully applied to different quantum states \cite{zhang2021detection}.



Among the investigation of quantum autoencoders, the compression rate is a significant indicator on the efficiency of quantum autoencoders. A given set of states may only admit a certain compression rate, which is closely related
with the inner pattern or the structure of input states \cite{romero2017quantum}. For example, the input states of qutrits in \cite{pepper2019experimental} were specifically constructed from polarization qubits using a single-photon source to allow them to be compressed to qubits with perfect compression rate of $100\%$. It is highly desirable to figure out general criteria for the compression rate of a quantum autoencoder. In this work, the compression rate is defined as the fidelity between the trash state (which represents the superfluous information and can be obtained via partially observing the original state) and the reference state (a fixed pure state for evaluation). We analyze the relationship between the compression rate and the inner structure of the input states. In particular, we establish a formula about the optimal compression rate for a set of known input states and provide an analytical solution to the optimal unitary operation that achieves the optimal compression rate.



For unknown input states, machine learning algorithms can be utilized to search for control fields with an optimal compression rate. In recent years, the combination of machine learning and quantum information has become a blooming area \cite{biamonte2017quantum}. It mainly aims at either employing machine learning techniques to aid quantum information tasks such as Hamiltonian learning \cite{wang2017experimental}, preparation of quantum states \cite{chen2014sampling}, and quantum gate control
\cite{dong2016learning}, \cite{niu2019universal}, \cite{wu2019learning} or utilizing quantum laws to speed up classical machine learning tasks \cite{dong2008quantum}, \cite{wu2020end}, \cite{li2020quantum}. Here, the task of implementing quantum autoencoders belongs to the former category, where different machine learning algorithms can be utilized for training the quantum autoencoders. Gradient methods have been used both theoretically \cite{wan2017quantum} and experimentally \cite{pepper2019experimental}, \cite{huang2020experimental}, and genetic algorithm (GA) has also been employed in optimizing a quantum autoencoder \cite{lamata2018quantum}, \cite{ding2019experimental}.

In this paper, we employ a closed-loop learning control approach \cite{rabitz2000whither}, \cite{chen2013closed}, \cite{wu2018data} to search for the control fields that can encode all the input states into the target latent space with a high compression rate. In particular, four different machine learning algorithms are applied in the proposed approach, with the purpose of training a quantum autoencoder with high efficiency. In addition, we present experimental results on compressing two 2-qubit states into two 1-qubit states using gradient methods to demonstrate the practical applicability of the theoretical and numerical results.

The main contributions of this work are summarized as follows:\\
$\bullet$ A necessary and sufficient condition is presented for perfect quantum autoencoders by analyzing the inner patterns of the input states, which further extends the result on the sufficient condition in \cite{huang2020experimental}.\\
$\bullet$ An upper bound of the compression rate for quantum autoencoders is proven to be the maximum sum of partial eigenvalues of the density operator representation of the input states. An analytic solution to the optimal unitaries that can achieve the upper bound is provided. These findings lay a foundation for theoretical research on quantum autoencoders. \\
$\bullet$ A closed-loop learning control framework is presented for searching for the control fields to achieve an optimal compression rate, and the training performance for quantum autoencoders using four learning algorithms including gradient algorithm, genetic algorithm, differential evolution, and evolutionary strategy is compared.\\
$\bullet$ A quantum optical experiment on compressing two 2-qubit states into two 1-qubit states is implemented to illustrate the practical applicability of the theoretical and numerical results. The effectiveness of training quantum autoencoders using learning algorithms may provide guidance for applying quantum autoencoders to different quantum applications. 

The structure of this paper is as follows. Several basic concepts and the research problems for quantum
autoencoders are introduced in Section \ref{sec:preliminary}. In Section \ref{sec:main}, theoretical results on the compression rate of a quantum autoencoder are established. Section \ref{sec:method} presents a closed-loop learning control approach for training a quantum autoencoder.
Numerical results on training quantum autoencoders using different learning algorithms are summarized in Section \ref{sec:numerical}. Experimental implementations of  quantum optical autoencoders are presented in Section \ref{sec:experiments}.
Conclusions are included in Section \ref{sec:conclusion}.

\emph{Notation}:
$a^*$ denotes the conjugate of $a$; $A^{T}$ is the transpose of $A$; $A^{\dagger}$ is the
conjugate and transpose of $A$; $A_{pq}$ is the $p$th row $q$th column element of $A$,
 $I$ is the identity matrix (dimension omitted if without ambiguity); $\textup{Tr}(A)$ is the trace of $A$. $\textup{rank}(A)$ is the rank of $A$.
 $\mathbb{H}$ denotes a Hilbert space, and $\mathbb{H}-\mathbb{S}$ denotes the subspace spanned by the vectors that are in $\mathbb{H}$, but not in $\mathbb{S}$, $\textup{dim}(\mathbb{H})$ denotes the dimension of $\mathbb{H}$; $\mathbb{C}$ is the sets of all complex numbers;
$|\psi\rangle$ is the unit complex vector representing a quantum (pure)
state and $\langle \psi |=(|\psi\rangle)^{\dagger}$; $|\cdot|$ means the norm of a vector; $\rho$ is the density matrix representing a quantum state; $\langle a | b \rangle$ denotes the inner product
of $a$ and $b$ with $\langle a | b \rangle=a^{\dagger}b$; $\mathbb{C}^d$ is the set of all $d$-dimensional
complex vectors; $\mathbb{U}^d$ is the set of all $d$-dimensional unitary operators, i.e., any $U\in\mathbb{U}^n$ satisfies $UU^{\dagger}=U^{\dagger}U=I$; $i$ as a subscript means an integer index; \rm{i} means imaginary unit, i.e., $\rm{i} =\sqrt{-1}$.
$\sigma_{x}, \sigma_{y}, \sigma_{z}$ are Pauli operators. $F(a,b)$ means the fidelity between state $a$ and state $b$. $a \otimes b$ denotes the tensor product between $a$ and $b$ and
$(\cdot)^{\otimes k}$ means that $(\cdot)$ tensored with itself $k$ times. $\lceil x \rceil$ takes the smallest integer that is not smaller than $x$.



\section{Preliminaries and research problems}\label{sec:preliminary}

\subsection{Quantum system}

The state of a finite-dimensional closed quantum system can be represented by a unit complex vector $|{\psi}\rangle$  and its dynamics can be described by the Schr\"{o}dinger equation:
\begin{equation}
\frac{d}{dt}|{\psi}(t)\rangle=-\frac{\rm{i}}{\hbar}H(t)|{\psi(t)}\rangle,
\label{eq:schron}
\end{equation}
where $\hbar$ is the reduced Planck constant (hereafter, we set $\hbar=1$ and use atomic unit) and $H(t)$ is the system Hamiltonian. $|\psi\rangle$ is actually a pure state, and a probabilistic mixture of pure states $\{p_k,|\psi_k\rangle\}$ is called a mixed state, which is described by a density matrix $\rho$. For pure states, $\rho=|\psi\rangle \langle \psi|$ and $\textup{Tr}(\rho^2)=1$. While in the general case $\rho=\sum_{k}p_k|\psi_k\rangle \langle \psi_k |$ with $p_k > 0 $ and $\sum_{k=1}p_k=1$, it is a Hermitian, positive semidefinite matrix satisfying $\textup{Tr}(\rho)=1$ and $\textup{Tr}(\rho^2)\leq 1$.


When we use control fields $\{u_j(t)\}_{j=1}^{M}$ to control the system, its system Hamiltonian in (\ref{eq:schron}) can be divided into two parts, i.e., $H(t)=H_0+H_c(t)=H_0+\sum_{j=1}^{M}u_j(t)H_j$, where $H_0$ is the time-independent free Hamiltonian of the system, $H_c(t)$ is the control
Hamiltonian representing the interaction of the system with the control
fields. For such a control system, its solution is given as $|\psi(t)\rangle = U(t)|\psi_0\rangle$ with $U(0)=I$, where the propagator $U(t)$ is formulated as follows:
\begin{equation}
\frac{d}{dt}U(t)=-{\rm{i}} [H_0+\sum_{j=1}^{M}u_j(t)H_j]U(t).
\label{eq: unitary propagator}
\end{equation}



\subsection{Hilbert space and composite system}
The vector space of closed quantum systems is also known as a Hilbert space. For each Hilbert space $\mathbb{H}=\mathbb{C}^n$, there exists a set of linearly independent vectors, forming a \emph{basis} for $\mathbb{H}$. The number of elements in the \emph{basis} is the \emph{dimension} of $\mathbb{H}$, namely $\textup{dim}(\mathbb{H})=n$. Let $\mathbb{H}_A$ and $\mathbb{H}_B$ be the state spaces of two quantum systems $A$ and $B$. Then the space of the composite system can be described by the tensor product of the spaces of its subsystems, that is $\mathbb{H}_{A}\otimes\mathbb{H}_B$. Partial trace is a map from the density matrix $\rho_{AB}$ on a composite space $\mathbb{H}_{A}\otimes\mathbb{H}_B$ onto a density matrix $\rho_A$ on a subspace $\mathbb{H}_A$. Let $\{|a_i\rangle\}$ be a basis of $\mathbb{H}_A$ and $\{|b_i\rangle\}$ be a basis of $\mathbb{H}_B$. The partial trace over system $B$ is defined as
\begin{equation}
\operatorname{Tr}_{B} (\rho_{AB})=\sum_{i} (I_{A} \otimes \langle b_i|) \rho_{AB} (I_A \otimes |b_i\rangle).
\end{equation}



\subsection{Problem formulation}
\begin{figure}[htb]
\centering
\begin{minipage}{0.45\linewidth}
  \centerline{\includegraphics[width=3.25in]{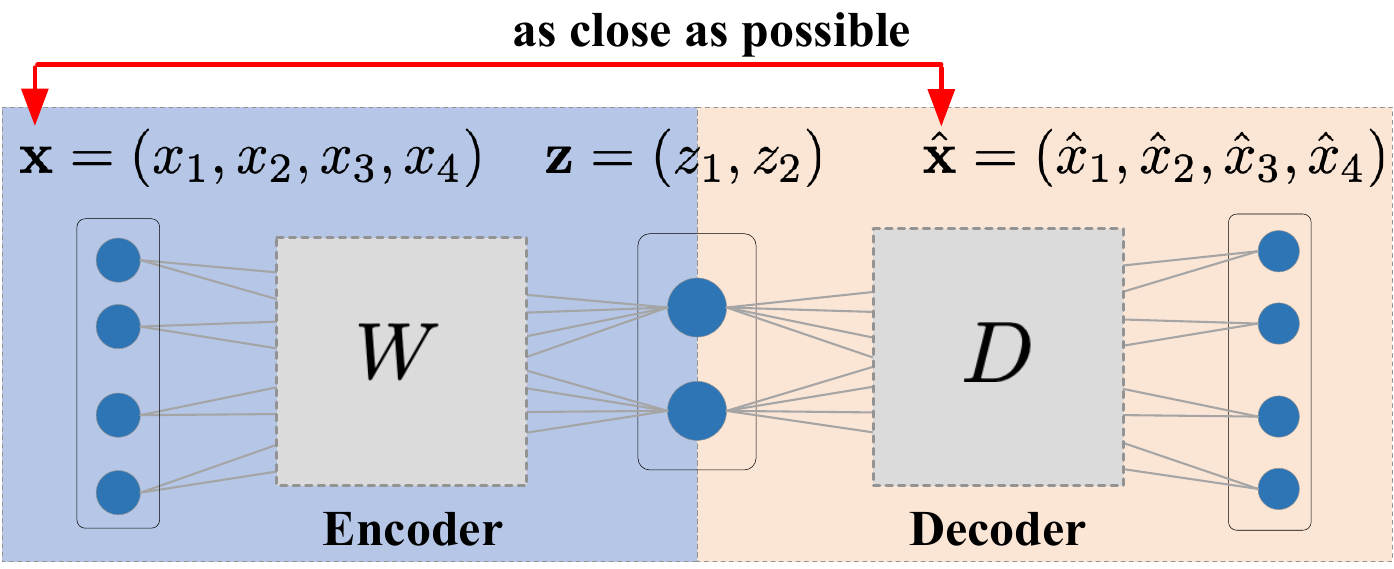}}
  \centerline{ (a)}
\end{minipage}
\vfill
\begin{minipage}{0.45\linewidth}
  \centerline{\includegraphics[width=3.25in]{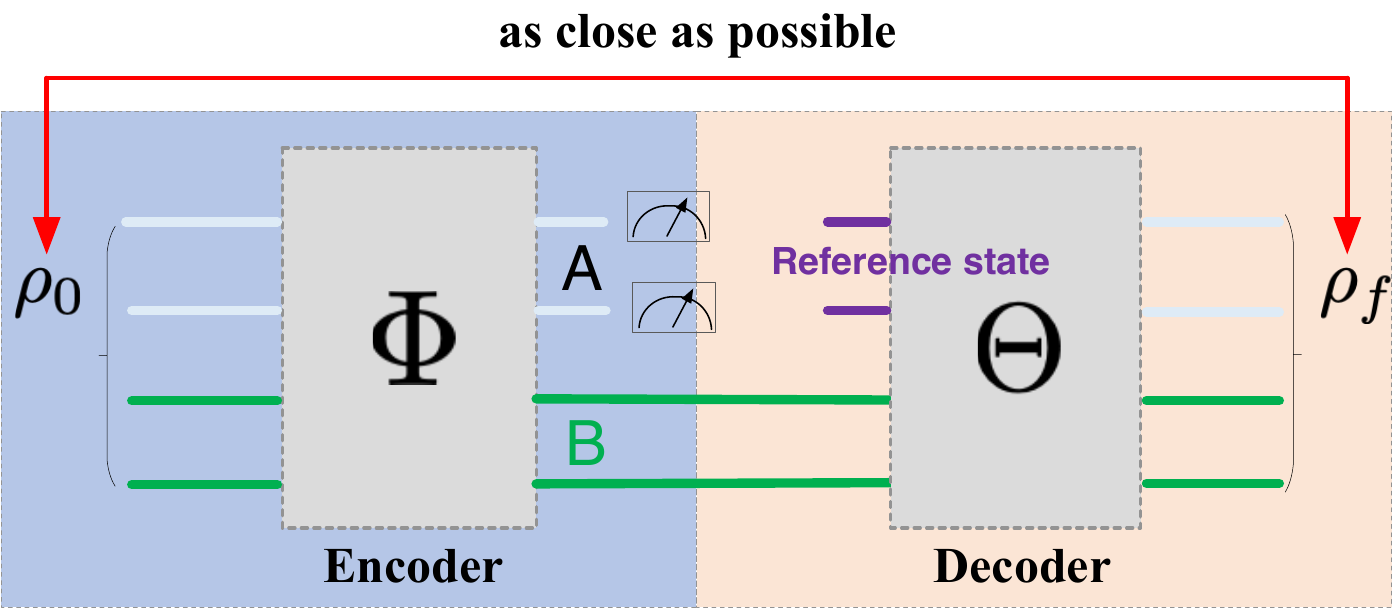}}
  \centerline{ (b)}
\end{minipage}
\caption{(a) Neural network representation of a classical 4-2-4 autoencoder. (b) Quantum
circuit representation of a quantum 4-2-4 autoencoder.}
\label{fig:classical-quantum autoencoder}
\end{figure}

A classical autoencoder is defined by a process of representing $n$-bit input string $\textbf{x}$ in corresponding $d$-bit ($0<d<n$) string $\textbf{z}$, as well
as recovering the original information from $\textbf{z}$, ending up in
$\hat{\textbf{x}}$. Tuples in the form of ``$n-d-n$" is often used to denote the bit number of the states $\textbf{x}$, $\textbf{z}$ and $\hat{\textbf{x}}$, respectively. Usually, the remaining $d$-bit is a compressed encoding form (with its space referred to as the latent
space or the bottleneck) of the string $\textbf{x}$. For example, in Fig. \ref{fig:classical-quantum autoencoder}(a), the map $W$ encodes a 4-bit input
$\textbf{x}=(x_1,x_2,x_3,x_4)$ into a 2-bit bottleneck state
$\textbf{z}=(z_1,z_2)$, after which the decoder $D$ attempts to reconstruct
the information, ending up with 4-bit output state
$\hat{\textbf{x}}=(\hat{x}_1,\hat{x}_2,\hat{x}_3,\hat{x}_4)$. In practical scenarios, a set of data, i.e., many items of $\textbf{x}$ are fed into the same neural networks to update the parameters for $W$ and $D$. To achieve this, a cost (loss) function, which evaluates the similarity between the original and the
reconstructed strings, is optimized according to the given set of input data.

A quantum autoencoder aims at encoding $n$-qubit state $\rho_0$ into
$d$-qubit state $\rho_B$, and recovering to $n$-qubit state $\rho_f$. For example, a graphical representation for a 4-2-4 quantum autoencoder is described in Fig. \ref{fig:classical-quantum autoencoder}(b), where $n=4$ and $d=2$.
The network includes two parts: 1) Encoder $\Phi$ (generally taken as a unitary transformation) reorganizes the 4-qubit input state $\rho_0$ onto the inner layer of the latent qubits, followed by discarding superfluous information contained in some of the input nodes. For example, this can be realized by tracing out the qubits representing these nodes;
2) Decoder $\Theta$ (another unitary transformation) reconstructs the 4-qubit state $\rho_f$ by using the combinations of the latent state and ancillary fresh qubits (initialized to the reference state). Likewise, the goal of the quantum autoencoder is to maximize the overlap between the recovered state and the original state. In this work, we mainly consider input pure states and unitary transformation maps for $\Phi$ and $\Theta$.



Different from a classical autoencoder, where both the encoding and decoding operations are usually performed to optimize the parameters for $W$ and $D$, the training of a quantum autoencoder can be reduced to optimizing the encoding transformation $\Phi$. After encoding quantum states from the data set using a trained unitary $\Phi$, quantum states can be naturally decoded by acting with $\Theta=\Phi^{\dagger}$ \cite{pepper2019experimental}. To train the quantum autoencoder, a practical way is to feed a set of input states into the network. As such, the task of a quantum autoencoder is to design an encoding map $\Phi$ to compress a given set of quantum states to optimize the compression rate, which can be defined as the fidelity between the original state $\rho_0$ and the recovered state $\rho_f$. In particular, we concentrate on the following questions: (i) under what condition can a quantum autoencoder perfectly compress a set of input states, i.e., having the compression rate 1? We call such an autoencoder as a perfect quantum autoencoder. (ii) how to train a quantum autoencoder towards the optimal compression rate using effective methods, such as machine learning? (iii) how high can the performance be achieved in the experimental realization of a quantum autoencoder?



\section{Theoretical results on compression rate}\label{sec:main}

In this section, we first investigate the compression rate of a quantum autoencoder, and employ
the fidelity between the trash states and the reference states as the cost function. In this work, the reference state is a fixed pure state and does not depend on the original state to be encoded. Then, we establish the necessary condition for a perfect quantum autoencoder by analyzing the inner patterns of the input states. Finally, the upper bound of the compression rate is investigated.

\subsection{Fidelity as compression rate}
A quantum autoencoder aims at encoding $n$ qubits into $d$ qubits (latent qubits), with $(n-d)$ qubits as trash qubits. Define the trash qubits as subsystem $A$ and the latent qubits as subsystem $B$, respectively. Let $\mathbb{H}_{AB}$, $\mathbb{H}_{A}$ and $\mathbb{H}_{B}$ denote the Hilbert spaces of the original space, the trash space and the latent space, with corresponding dimensions $N$, $N_A$ and $N_B$, respectively. We have $\mathbb{H}_{AB}=\mathbb{H}_A\otimes \mathbb{H}_B$ and $N=N_AN_B$ \cite{nielsen2010quantum}.

Denote the reference state as $\rho_{\textup{ref}}=|\Psi_{\textup{ref}}\rangle \langle \Psi_{\textup{ref}}|$, where $|\Psi_{\textup{ref}}\rangle$ is an arbitrary fixed pure state. For an input state $\rho_0=|\psi_0\rangle \langle \psi_0|$, considering an encoding map $U$, we have $\rho_T=|\psi_T\rangle \langle \psi_T|=U\rho_0 U^{\dagger}$. By means of partial trace, the states of the trash space and the latent space are $\rho_{A} = \textup{Tr}_{B}(\rho_T)$ and $\rho_{B} = \textup{Tr}_{A}(\rho_T)$, respectively. Then, fresh qubits are initialized and a decoding transformation $U^{\dagger}$ is performed to obtain the recovered state $\rho_f = U^{\dagger} (\rho_{\textup{ref}}\otimes \rho_{\textup{B}})  U$. To evaluate the efficiency of a quantum autoencoder, it is natural to use the fidelity between the recovered
state and the initial state as an objective function
\begin{equation}
\begin{split}
J_1 = &F(\rho_0,\rho_f)=\langle \psi_0| U^{\dagger} \rho_{\textup{ref}} \otimes \textup{Tr}_{A}( U \rho_0 U^{\dagger}) U |\psi_0\rangle.
\label{eq:J_recover}
\end{split}
\end{equation}

Let $|\psi_B\rangle \in \mathbb{H}_B$ be the latent state in the latent space and $|\Psi_{\textup{ref}}\rangle \in \mathbb{H}_{A}$ be the reference state in the trash space. When a unitary perfectly disentangles the input state into two parts as
\begin{equation}
U |\psi_{0}\rangle = |\psi_{A}\rangle \otimes |\psi_B\rangle=|\Psi_{\textup{ref}}\rangle \otimes |\psi_B\rangle,
\label{eq:disentangle}
\end{equation}
we have $ U \rho_0 U^{\dagger}=\rho_{\textup{ref}} \otimes |\psi_B\rangle  \langle \psi_B|$. The recovered state reads $\rho_f=U^{\dagger} [\rho_{\textup{ref}} \otimes \textup{Tr}_{A}( U \rho_0 U^{\dagger})] U=
U^{\dagger}[\rho_{\textup{ref}} \otimes |\psi_B\rangle  \langle \psi_B| ]U=|\psi_0\rangle \langle \psi_0|$.
Hence, $J_1=F(\rho_0,\rho_f)=1$ can be achieved. Here, the overlap between the trash state and the reference state can be used to quantify the efficiency of a quantum autoencoder, leading to another objective function
\begin{equation}
J_2 =F(\rho_{\textup{ref}},\rho_A )= \langle \Psi_{\textup{ref}}| \textup{Tr}_{B}(U \rho_0 U^{\dagger}) |\Psi_{\textup{ref}}\rangle.
\label{eq:J_trash}
\end{equation}

Generally, $J_1$ and $J_2$ are not identical; in fact, $J_1 \leq J_2 $ \cite{verdon2018universal}. Combined with the fact that the perfect fidelity of $J_1$ achieves if and only if perfect disentangling is realized \cite{romero2017quantum}, we draw the the following conclusion (see \ref{app:disentanglement} for the proof).
\begin{lem}
The optimal value of $J_1=J_2=1$ can be achieved if and only if $U$ systematically
\emph{disentangles} the high-dimensional input state $|\psi_0\rangle$ into two low-dimensional parts, with the trash state set as a fixed reference state i.e., $
U |\psi_{0}\rangle = |\psi_{A}\rangle \otimes |\psi_B\rangle=|\Psi_{\textup{ref}}\rangle \otimes |\psi_B\rangle$.
\label{pro:disentanglement}
\end{lem}

In this paper, we use $J_2$ to characterize the performance of a quantum autoencoder considering the facts that
(i) $J_1$ and $J_2$ are closely related with each other and can achieve the same value $1$ for a perfect quantum autoencoder simultaneously;
and (ii) $J_2$ can be measured within a smaller trash space involving less resource and has no requirement to access the input states.

Now, we consider the compression of a set of input states using the same unitary operator $U$. Let $\{p_k,|{\psi_0^{k}}\rangle\}_{k=1}^{Q}$ be an ensemble of pure input states, and we may write the corresponding density operator as follows:
\begin{equation}
\rho=\sum_{k=1}^{Q}p_k|{\psi_0^{k}}\rangle \langle {\psi_0^{k}}|.
\label{eq:composed rho}
\end{equation}
For each input state $|{\psi_0^k}\rangle$, the performance is evaluated by
\begin{equation}
J(U,|\psi_0^k\rangle)=\langle \Psi_{\textup{ref}}| \textup{Tr}_{B}(U
|\psi_0^{k}\rangle \langle \psi_0^{k}| U^{\dagger})  |\Psi_{\textup{ref}}\rangle,
\label{eq:single equation}
\end{equation}
averaging over which the overall objective function can be written as
\begin{equation}
\begin{split}
J(U)&=\sum_{k=1}^{Q} p_k \langle \Psi_{\textup{ref}}| \textup{Tr}_{B}(U |\psi_0^{k}\rangle
\langle \psi_0^{k}| U^{\dagger}|)   |\Psi_{\textup{ref}}\rangle  \\
  & = \langle \Psi_{\textup{ref}}| \textup{Tr}_{B}
  (\sum_{k=1}^{Q} p_k U |\psi_0^{k}\rangle \langle \psi_0^{k}|
  U^{\dagger} )      |\Psi_{\textup{ref}}\rangle \\
  & =  \langle \Psi_{\textup{ref}}| \textup{Tr}_{B} (U \rho
  U^{\dagger})  |\Psi_{\textup{ref}}\rangle.
\end{split}
\label{eq:overall function}
\end{equation}
Finally, we have the fidelity between $|\Psi_{\textup{ref}}\rangle$ and the trash state $\textup{Tr}_{B} (U \rho
  U^{\dagger})$ as the objective function for a quantum autoencoder, which is defined as the compression rate in this
  work. Theoretically, $|\Psi_{\textup{ref}}\rangle$ can be any pure state, but it is usually set as a pure state that is easy to generate in physical implementation, such as $|0\rangle ^ {\otimes (n-d)}$.


\subsection{The inner patterns of input states for perfect compression rate}\label{subsec:perfect}
Consider a set of states, with support on a subspace $\mathbb{S} \subset \mathbb{H}$. Traditionally, $\lceil \log_{2}\textup{dim}(\mathbb{H})\rceil$ qubits are needed to represent the states in $\mathbb{S}$. As such,  quantum autoencoders aim at determining an encoding scheme that employs only $\lceil \log _{2}\textup{dim}(\mathbb{S})\rceil$ qubits to represent the states \cite{romero2017quantum}.The possibility of rearranging them into a latent form lies in the symmetries among the inner structure of the input states. For example, this encoding scenario is usually associated with eigenstates of
many-body systems due to special symmetries \cite{romero2017quantum}. Moreover, the maximum compression rate of a quantum autoencoder is usually limited by the inner structure of the input data set. Consequently, a given set of states might only admit a small or null compression rate. Here, we first concentrate on the case of a perfect compression rate $1$ (also called a perfect autoencoder).



In this work, the dimension of the subspace spanned by a set of vectors is termed as $\textup{SpanF}(\cdot)$ to distinguish from the length of each vector. It is clear that $\textup{SpanF}(\cdot)$ is actually the number of maximal linearly independent vectors from the vector set.
Define $\Omega_V:=\{|v_1\rangle,|v_2\rangle,\ldots,|v_{Q}\rangle\}$, with $|v_k\rangle \in \mathbb{C}^{N}$ represented by a column vector $[v_{1k},v_{2k},\dots,v_{Nk}]^{{T}}$. Then these $Q$ vectors compose a matrix $V=[v_{jk}]$. We have $\textup{SpanF}(\Omega_V) \equiv  \textup{rank}(V)$ and the following conclusion.

\begin{lem}
	If a quantum autoencoder can compress a set of input states
	$\{|\psi_0^k\rangle\}_{k=1}^{Q}$ with perfect compression rate using the same unitary $U$, then it can compress linear combination of the input states $|\psi_0^{\vec{p}}\rangle=\sum_{k=1}^{Q}p_k|\psi_0^k\rangle$ ($p_k\in \mathbb{C}$ is normalized to guarantee $||\psi_0^{\vec{p}}\rangle|=1$) with perfect compression rate.
	\label{th:perfect autoencoder}
\end{lem}

\begin{pf}
According to (\ref{eq:disentangle}), a perfect quantum autoencoder disentangles the initial states, with the trash states set as the reference state, i.e., $
U|\psi_0^k\rangle = |\Psi_{\textup{ref}}\rangle \otimes |\varphi_B^k\rangle, k=1,2,...,Q,$ where $|\varphi_B^k\rangle \in \mathbb{H}_{B}$. Then, we have the following equation:
\begin{equation}
U|\psi_0^{\vec{p}}\rangle=\sum_{k=1}^{Q}p_k U |\psi_0^k\rangle=\sum_{k=1}^{Q}p_k |\Psi_{\textup{ref}}\rangle
\otimes |\varphi_B^k\rangle=|\Psi_{\textup{ref}}\rangle \otimes |\varphi_B^{\vec{p}}\rangle,
\end{equation}
where $|\varphi_B^{\vec{p}}\rangle=\sum_{k=1}^{Q}p_k |\varphi_B^k\rangle$ is a
complex vector in $\mathbb{H}_B$.
In addition, since $U|\psi_0^{\vec{p}}\rangle=|\Psi_{\textup{ref}}\rangle
\otimes |\varphi_B^{\vec{p}}\rangle=|\psi^{\prime}\rangle$ is a unit complex
vector in $\mathbb{H}_{AB}$, we have $\textup{Tr}(|\psi^{\prime}\rangle \langle
\psi^{\prime}|)=1$ and $\textup{Tr}_A(|\psi^{\prime}\rangle \langle
\psi^{\prime}|)=\textup{Tr}_A(|\Psi_{\textup{ref}}\rangle
\otimes |\varphi_B^{\vec{p}}\rangle \langle \Psi_{\textup{ref}}|
\otimes \langle \varphi_B^{\vec{p}}|)=|\varphi_B^{\vec{p}}\rangle  \langle
\varphi_B^{\vec{p}}|$. As partial trace is a trace preserving operation, we
have $\textup{Tr}(|\varphi_B^{\vec{p}}\rangle  \langle
\varphi_B^{\vec{p}}|)=\textup{Tr}(|\psi^{\prime}\rangle \langle
\psi^{\prime}|=1$. Hence $|\varphi_B^{\vec{p}}\rangle \in \mathbb{H}_B$ is a unit complex vector.
Therefore, the quantum autoencoder compresses the state
$|\psi_0^{\vec{p}}\rangle$ into the latent state $|\varphi_B^{\vec{p}}\rangle$ by unitary $U$.
\hfill $\Box$
\end{pf}

According to Lemma \ref{th:perfect autoencoder}, a quantum autoencoder that can perfectly encode a set of input states with perfect fidelity can also compress their linear combinations with perfect fidelity. Hence, a perfect autoencoder might be restricted by the number of linearly independent vectors among the input states.  The dimension of the target latent space puts a restriction on the number of maximal linearly independent vectors among the latent states, and further constrains the number of maximal linearly indenpedent vectors from the input states. In particular, we have the following proposition:


\begin{prop}
	A necessary condition for a perfect quantum autoencoder is that the maximal number of linearly
	independent vectors among the input states is not more than the dimension of the target latent space.
	\label{lm:necessary}
\end{prop}

\begin{pf}

Firstly, performing a unitary transformation on the vectors in $\Omega_V$, we have the output vectors as $
|v^{\prime}_k\rangle=U|v_k\rangle,k=1,\ldots,Q$.
Denote $\Omega_{V^{\prime}}=\{|v_1^{\prime}\rangle,...,|v_k^{\prime}\rangle,\ldots,|v_Q^{\prime}\rangle\}$ and the matrix notation of $\Omega_{V^{\prime}}$ as $V^{\prime}$.
It is clear that $V^{\prime}=UV$. Since $\textup{rank}(V^{\prime})=\textup{rank}(V)$, we have $\textup{SpanF}(\Omega_{V^{\prime}})=\textup{SpanF}(\Omega_V)$.
Then, the input states are divided into the trash states and the latent states, which means that the composite space is divided into the trash space $\mathbb{H}_A$ and the latent space $\mathbb{H}_B$. Let $\Omega_{V^{\prime}_B}$ be the subspace spanned by the latent states. We have
$\textup{SpanF}(\Omega_{V^{\prime}_B}) \leq \textup{dim}(\mathbb{H}_{B})$, where $\textup{dim}(\mathbb{H}_{B})$ represents the dimension of the target latent space, e.g., $\textup{dim}(\mathbb{H}_{B})=2^{m}$ for $m$ latent qubits. Since the trash state is fixed as $|\Psi_{\textup{ref}}\rangle$ for a perfect autoencoder, the dimension of the trash space is 1. Then we obtain $\textup{SpanF}(\Omega_{V})= \textup{SpanF}(\Omega_{V^{\prime}})=\textup{SpanF}(\Omega_{V^{\prime}_B}) \leq \textup{dim}(\mathbb{H}_{B})$.
That is to say, for a perfect quantum autoencoder, the dimension of the input space spanned by the input states should be no greater than the dimension of the target latent space.
\hfill $\Box$
\end{pf}


\subsection{The upper bound of the compression rate for given input states}\label{subsec:bound}
Now we focus on the upper bound of the compression rate defined by the fidelity between the
reference state $|\Psi_{\textup{ref}}\rangle$ and the trash state $\textup{Tr}_{B} (U \rho U^{\dagger})$.
For this purpose, we need to find a unitary $U\in\mathbb{U}^N$ that maximizes $\langle \Psi_{\textup{ref}}| \textup{Tr}_{B} (U \rho
  U^{\dagger})  |\Psi_{\textup{ref}}\rangle$.

\begin{lem}\cite{jacobs2014quantum}
Among all unitaries $U\in \mathbb{U}^N$, the highest fidelity between a given pure state $|\psi\rangle$ and a mixed state $\rho^U=U\rho U^{\dagger}$ with $\rho$ a given state is only determined by the maximal eigenvalue of $\rho$: $\max_{U\in\mathbb{U}^N}F(|\psi\rangle,U\rho U^{\dagger})=\max_i \lambda_i(\rho)$.
	\label{pro:maximum fidelity}
\end{lem}

According to Lemma \ref{pro:maximum fidelity} (see \ref{app:maximum fidelity} for the proof) and  $F(|\psi\rangle,U^{\dagger}\rho U)=F(\hat{U}|\psi\rangle, \hat{U} U^{\dagger}\rho U \hat{U}^{\dagger})$, a unitary transformation $\hat{U}$ on $|\psi\rangle$ is equivalent to the inverse transformation $\hat{U}^{\dagger}$ on $U\rho U^{\dagger}$. Besides, a unitary transformation on a density operator does not change its eigenvalues. Hence, the choice of $|\psi\rangle$ does not influence the maximum fidelity between $|\psi\rangle$ and $\rho^U=U\rho U^{\dagger}$. To measure  $J(U)=F(|\Psi_{\textup{ref}}\rangle,\textup{Tr}_B(U\rho U^{\dagger})$, we set the reference state as a benchmark state, e.g., $|\Psi_{\textup{ref}}\rangle=[1,0,...,0,...]^{T}$. Now, $J(U)$ can be written as the first diagonal element of $\textup{Tr}_{B}(U \rho U^{\dagger})$. According to \ref{app:partialtrace}, the diagonal elements of a reduced state are actually partial sums of $n$ diagonal elements of the state $U \rho U^{\dagger}$, where $n$ equals the dimension of subsystem $B$.
Hence, the maximum fidelity $J(U)$ for a quantum autoencoder amounts to the maximum partial sum of $N_B$ diagonal elements. Denote $\textup{SD}_n(\cdot)$ as the sum of the first $n$ diagonal elements of a matrix and $\lambda_i(\cdot)$ as the $i$-th eigenvalue of a positive semi-definite operator (in descending order). We introduce the following theorem:
\begin{thm}
Given a density operator $\rho$, the maximum sum of $n$ diagonal elements of $\rho^U=U\rho U^{\dagger}$ with $U\in \mathbb{U}^N$ equals the sum of the first $n$ eigenvalues (descending order) of $\rho$, i.e., $\max_{U\in\mathbb{U}^N} \textup{SD}_n(U\rho U^{\dagger})=\sum_{i=1}^{n}\lambda_i(\rho)$.
\label{th:eigen-diagnonal}
\end{thm}
\begin{pf}
It is clear that for $n=N$, $\textup{SD}_n(\rho^U)=\sum_{i=1}^{n}[\rho^{U}]_{ii}=\textup{Tr}(\rho^{U})=\textup{Tr}(\rho)=\sum_{i=1}^{N}\lambda_i(\rho)$.
For $n<N$, we employ the matrix differential method to determine the maximum value of $\textup{SD}_n(\rho^U)$.
Denote an operator $V=\sum_{i=1}^{n}|i\rangle \langle i | $, with $|i\rangle$ as the $i$-th column of $I_N$ and we have
$\textup{SD}_n(\rho^{U})=\sum_{i=1}^{n} \langle i | U \rho U^{\dagger} |i\rangle = \sum_{i=1}^{n} \textup{Tr} ( U \rho U^{\dagger} |i\rangle \langle i |) = \textup{Tr}( U \rho U^{\dagger} V)$.
Since $\textup{Tr}(AB)=\textup{Tr}(BA)$, we have $\textup{SD}_n(\rho^{U})=\textup{Tr}(\rho U^{\dagger}V U)$.

Then, we use matrix differentiation \cite{bhatia1997matrix}, \cite{zhang2017matrix} to figure out the maximum value of $\textup{Tr}(\rho U^{\dagger}V U)$.
Let us introduce a Lagrange function as follows:
\begin{equation}
L:=\textup{Tr}(\rho U^{\dagger}VU)+\textup{Tr}(\Lambda +\Lambda^{\dagger})(UU^{\dagger}-I),
\end{equation}
where $\Lambda$ is the Lagrange multiplier matrix. The derivation of $L$ against $U^{*}$ is written as
\begin{equation}
\frac{ \partial L}{\partial U^{* }}=VU\rho+(\Lambda+\Lambda^{\dagger})U=0.
\end{equation}
Thus, we have $VU\rho=-(\Lambda+\Lambda^{\dagger})U$, and $VU\rho U^{\dagger}=-(\Lambda+\Lambda^{\dagger})$.
Since $(\Lambda+\Lambda^{\dagger})$ is Hermitian, $VU\rho U^{\dagger}$ must be Hermitian. Denote $V$ and $U\rho U^{\dagger}$ in block matrix representations as $$V=\left[\begin{array}{cc}{I_n} & 0\\ 0 & 0 \end{array}\right],\quad U\rho U^{\dagger}=\left[\begin{array}{cc}{H_{1}} & H_{2}\\ H_{3} & H_{4} \end{array}\right].$$
We have $$
VU\rho U^{\dagger}-U\rho U^{\dagger}V= \left[\begin{array}{cc}{0} & H_{2}\\ -H_{3} & 0 \end{array}\right].$$
To satisfy $VU\rho U^{\dagger}=U\rho U^{\dagger}V$, $H_{2}$ and $H_{3}$ must be zero matrixes, which means
$$U\rho U^{\dagger}=H=\left[\begin{array}{cc}{H_1} & 0\\ 0 & H_4 \end{array}\right].$$ A further calculation reveals that
\begin{equation}
VU\rho U^{\dagger}=
\left[\begin{array}{cc}{I_n} & 0\\ 0 & 0 \end{array}\right] \left[\begin{array}{cc}{H_1} & 0\\ 0 & H_4 \end{array}\right]
=\left[\begin{array}{cc}{H_1} & 0\\ 0 & 0 \end{array}\right].
\end{equation}
Hence, $\textup{SD}_n(\rho^{U})=\textup{Tr}(VU\rho U^{\dagger})=\textup{Tr}(H_1)=\sum_{i=1}^{n} \lambda_i(H_1)$.

Let the eigenvectors of $H_1$ be $\{x_1^i\}_{i=1}^n$ ($x_1^i$ corresponds to the eigenvector of $\lambda_i(H_1)$). We have $H_1 x_1^i =\lambda_i x_1^i$ for $i=1,2,...,n$.
It is clear that
\begin{equation}
\left[\begin{array}{cc}{H_1} & 0\\ 0 & H_4 \end{array}\right] \left[\begin{array}{c} x_1^i \\ 0 \end{array}\right]=\left[\begin{array}{c} H_1x_1^i \\ 0 \end{array}\right]=\lambda_i \left[\begin{array}{c} x_1^i \\ 0 \end{array}\right].
\end{equation}
Hence, the eigenvalues of $H$ consist of those of $H_1$ and $H_4$. Therefore, we have the following conclusion:
\begin{equation}
\max_{U\in\mathbb{U}^N} \textup{SD}_n(\rho^{U}) = \sum_{i=1}^{n} \lambda_i(H_1) = \max \sum_{i=1}^{n} \lambda_i(\rho).
\end{equation}
\hfill $\Box$
\end{pf}

It is worth mentioning that the conclusion in Theorem \ref{th:eigen-diagnonal} can also be proven using quantum control landscape theories \cite{wu2008characterization}.
According to Theorem \ref{th:eigen-diagnonal}, the maximum value of $J(U)$ equals the maximum value of the first one diagonal element of $\textup{Tr}_B(\rho^U)$, which is the sum of first $N_B$ maximal diagonal elements of $\rho^U$. That is:
\begin{equation}
\max_{U\in\mathbb{U}^N} J(U) = \max_{U\in\mathbb{U}^N} \textup{SD}_{N_B}(U\rho U^{\dagger}) = \sum_{i=1}^{N_B}\lambda_i(\rho).
\label{eq:eigens}
\end{equation}
The solution of $U$ that maximizes $J(U)$ has the general form of $$U \rho U^{\dagger}=\left[\begin{array}{cc}{H_1} & 0\\ 0 & H_4 \end{array}\right]=\left[\begin{array}{cc}{M_1D_1M^{\dagger}} & 0\\ 0 & M_4 D_4M_4^{\dagger}\end{array}\right],$$ where $M_1$ is an $N_B$ dimensional unitary operator, and $M_4$ is an $(N-N_B)$ unitary operator. $D_1$ and $D_4$ are diagonal operators consisiting of eigenvalues of $\rho$, i.e., $D_1 = \textup{Diag}(\lambda_1,\lambda_2,...,\lambda_{N_{B}})$ and  $D_4 = \textup{Diag}(\lambda_{N_{B}+1},...,\lambda_{N})$.

For the maximum value of compression rate in equation (16), we propose a lower bound depending on the latent space dimension $N_B$ and the rank of $\rho$ (denoted as $R$) as \begin{equation}
\max_{U \in \mathbb{U}_N} J(U)=\sum_{i=1}^{N_B} \lambda_i(\rho) \geq  \min(\frac{N_B}{R},1). 
\label{eq:lower bound}
\end{equation}
The proof of the lower bound (\ref{eq:lower bound}) is as follows:
Recalling that the eigenvalues of $\rho$ are in the descending order, we can write them as $(\lambda_1,....,\lambda_{R},0,...,0)$, where we have $\lambda_1 \geq \lambda_2 \geq ... \geq \lambda_R >0$ and $\sum_{i=1}^{R}\lambda_i=1$. When $R > N_B$,  since $\lambda_1 \geq \lambda_2 \geq ... \geq \lambda_R >0$, we have
\begin{equation} 
\frac{1}{N_B}\sum_{i=1}^{N_B}\lambda_i \geq   \frac{1}{R-N_B}\sum_{i=N_B+1}^{R}\lambda_i.
\end{equation} 
Clearly, $ \min \sum_{i=1}^{N_B} \lambda_i=\frac{N_B}{R}$ can be realized when $\lambda_1=\lambda_2=...=\lambda_{R}=\frac{1}{R}$. 
When $R \leq N_B$, we have $\sum_{i=1}^{N_B} \lambda_i(\rho)=\sum_{i=1}^R \lambda_i +\sum_{i=R+1}^{N_B} \lambda_i =\sum_{i=1}^R \lambda_i=1$. We combine the two cases together and obtain a general conclusion as $\sum_{i=1}^{N_B} \lambda_i(\rho) \geq \min(\frac{N_B}{R},1)$.

Now, we consider the general case with the reference state $|\Psi_{\textup{ref}}\rangle \neq [1,0,\ldots,0]^{T}$. According to \ref{app:unitary solution}, there exists $U_A$ that can transform any $|\Psi_{\textup{ref}}\rangle $ to the benchmark state, i.e., $U_A|\Psi_{\textup{ref}}\rangle=[1,0,\ldots,0]^{T}$. Hence, a more general form of the optimal unitary can be given as
\begin{equation}
U=U_A^{\dagger} \otimes I_B \left[\begin{array}{cc}{M_1} & 0\\ 0 & M_4 \end{array}\right]W
\end{equation}
with $ W \rho W^{\dagger}=\textup{Diag}(\lambda_1,\lambda_2,...,\lambda_N)$.
Finally, we have the following corollary:
\begin{cor}
The maximal value of $J(U)$ equals the sum of the first $N_B$ maximum eigenvalues of $\rho$.
The optimal solution is $U=U_A^{\dagger} \otimes I_B U_{AB}$, where
$U_A |\psi_{\textup{ref}}\rangle=[1,0,...,0]^{T}$,
 $U_{AB}=\left[\begin{array}{cc}{M_{1}} & 0\\ 0 & M_{4} \end{array}\right]W$, $M_{1}$ and $M_4$ are $N_{B}$ and $(N-N_{B})$ dimensional unitary operators, respectively, and $W$ diagonalizes $\rho$ as $W \rho W^{\dagger}=\textup{Diag}(\lambda_1,\lambda_2,...,\lambda_N)$.
 \label{cor:limit value}
\end{cor}

\begin{rem}
Although a unitary $U$ does not change the eigenvalues of the initial state $\rho$, the eigenvalues of the trash state $\rho_A=\textup{Tr}_{B}(U \rho U^{\dagger})$ and the latent state $\rho_{B}=\textup{Tr}_{A}(U \rho U^{\dagger})$ can both be influenced by $U$. Hence, the decomposition between the latent state and the trash state depends on both the eigenvalues of the state to be encoded and the unitary  performed on the states. Furthermore, Corollary 6 reveals that the best decomposition which achieves the optimal compression rate is solely determined by the eigenvalues of the state.
\end{rem}

\subsection{A special case}\label{subsec:cases}
Here, we investigate the case of perfect autoencoder. Recall the necessary condition for a perfect autoencoder in Proposition \ref{lm:necessary} is that the number of linearly independent vectors from the input states is not more than the dimension of the target latent space.  On the other hand, it has been found that when the number of linearly independent vectors from the input states is not more than the dimension of the target latent space, a perfect quantum autoencoder with compression rate 1 can be achieved  \cite{huang2020experimental}. Such a conclusion can also be deduced from Theorem \ref{th:eigen-diagnonal} and Corollary \ref{cor:limit value}. Hence, the necessary condition (the maximum number of linearly independent vectors among the input states is not more than the dimension of the latent space) is also sufficient for achieving a perfect quantum autoencoder, we have the following theorem:

\begin{thm}
A perfect quantum autoencoder can be achieved if and only if the number of linearly independent vectors among the input states is \textbf{not more than} the dimension of the target latent space.
\label{them:perfect}
\end{thm}


\section{Closed-loop learning control for quantum autoencoders}\label{sec:method}
	For a set of given input states,  an analytical solution to the optimal unitary operation that achieves the optimal compression rate has been established in Section 3. However, when the input states are unknown, the calculation of the optimal unitary operation can not be performed. This scenario will be the focus of this section, and we apply learning algorithms to search for the optimal control fields that maximize the compression rate. To compress a set of input states
$\{p_k,|{\psi}_{0}^{k}\rangle\}_{k=1}^{Q}$ using the same control strategy $\textbf{u}:=\{u_j(t)\}_{j=1}^{M}$, an average loss function regarding different input states is introduced $J(\textbf{u}):= \mathbb{E}[J(\textbf{u}, \rho)] = \sum_{k=1}^{Q}p_kJ(\textbf{u}, |\psi_0^{k}\rangle)$. The problem of
finding the unitary operator $\Phi(\textbf{u})$ for a quantum autoencoder can be formulated as
\begin{equation}
\begin{split}
\displaystyle  \ \ \  & \max_{\textbf{u}} \
J(\textbf{u}):=\max_{\textbf{u}} \mathbb{E}[J(\textbf{u}, \rho)] =\max_{\textbf{u}} \sum_{k=1}^{Q}p_kJ(\textbf{u}, |\psi_0^{k}\rangle)  \\
\text{s.t.} \\
&\left \{
\begin{split}
&\frac{d}{dt}\Phi(\textbf{u}(t))=-{\rm{i}}(H_0+\sum_{j=1}^{M}u_j(t)H_j)\Phi(\textbf{u}(t)), \quad  t \in [0, T] \\
&|\psi_T^k \rangle=\Phi(\textbf{u}(T))|{\psi}_{0}^{k}\rangle, \quad k=1,2,...,Q, \\
& J(\textbf{u}, |\psi_0^k\rangle)= \langle \Psi_{\textup{ref}}|
\textup{Tr}_{B} (|\psi_T^k \rangle \langle \psi_T^k|)\Psi_{\textup{ref}}\rangle.
\label{eq:problem}
\end{split}\right.\\
\end{split}
\end{equation}

\begin{figure}
\centering
\includegraphics[width=0.48\textwidth]{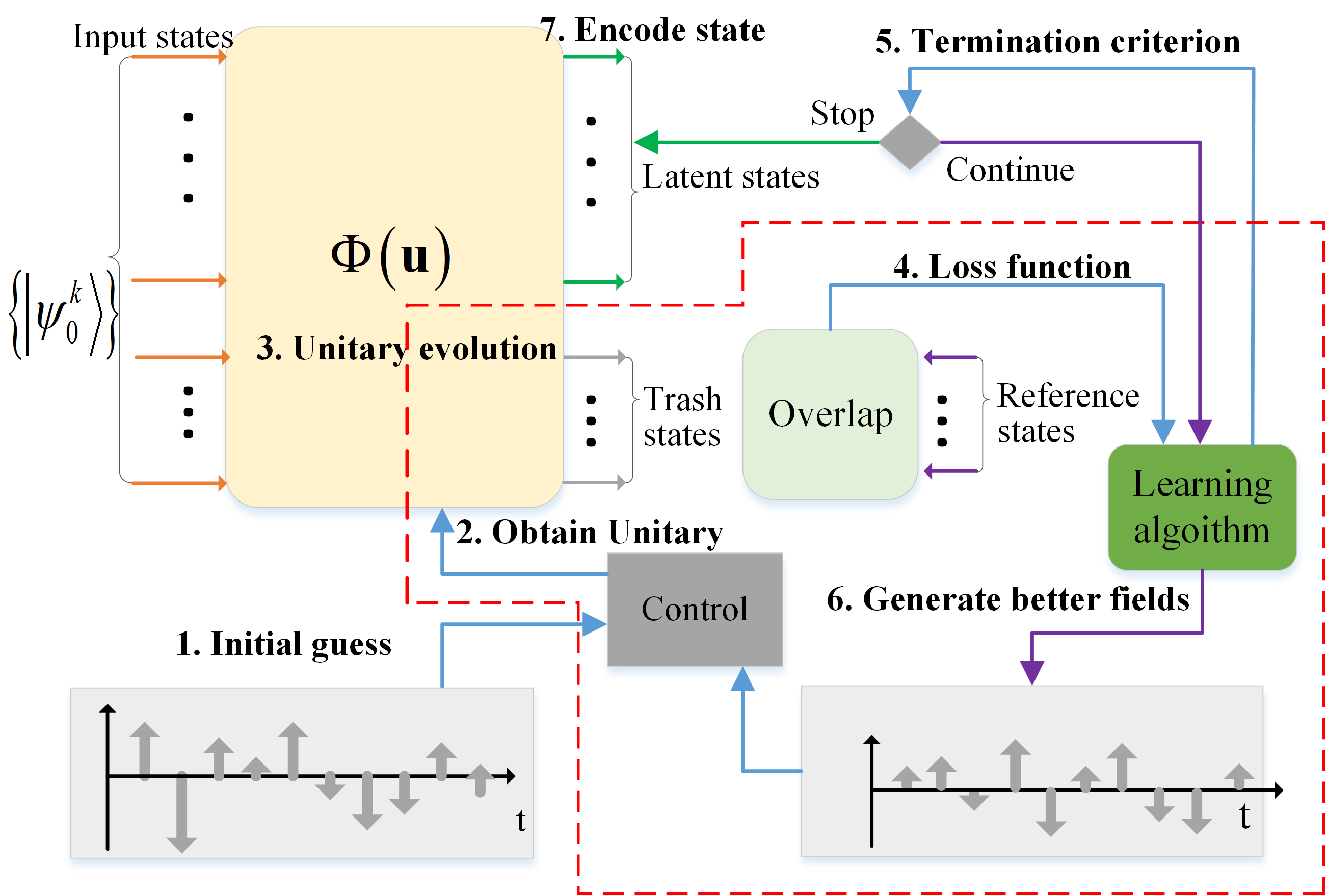}
\caption{The procedure of the closed-loop learning control approach for a quantum autoencoder.}
\label{fig:framework of quantum autoencoder}
\end{figure}

To find a numerical optimal control solution for problem (\ref{eq:problem}), we adopt a closed-loop
learning control approach as shown in Fig. \ref{fig:framework of quantum autoencoder} to train the quantum autoencoder. The approach starts from an initial guess, and employs learning algorithms to generate or suggest a better control strategy, based on the learning performance of the prior trial. For each trial, it is an open-loop
control process, while the control performance will be sent back to the learning algorithm to guide the optimization for the control strategy \cite{chen2013closed}. The general procedure is summarized as follows:\\
\textbf{Step 1.} Generate an initial guess of feasible control field $\textbf{u}$;\\
\textbf{Step 2.} Obtain the unitary operator $\Phi(\textbf{u})$ according to (\ref{eq: unitary propagator});\\
\textbf{Step 3.} Perform the unitary transformation $\Phi(\textbf{u})$ for all the input states and obtain their trash states;\\
\textbf{Step 4.} Compute the average objective function $J(\textbf{u})$;\\
\textbf{Step 5.} Check the termination criterion, if $J(\textbf{u})$ converges go to Step 7, otherwise go to Step 6;\\
\textbf{Step 6.} Suggest a better control field $\textbf{u}$ using machine learning algorithms and go to Step 2;\\
\textbf{Step 7.} Encode the input states by the optimal unitary operator and obtain the compressed form of the input states.

A crucial step of the general procedure lies in Step 6 that suggests better control fields. For example, gradient methods (GD) have gained wide popularity as a result of their effectiveness \cite{khaneja2005optimal}, \cite{chen2014sampling}. However, they usually rely on accurate system models for gradient evaluation, which may result in local traps. Stochastic searching methods such as evolutionary algorithms can step over local maxima and have been widely used in complex quantum control problems due to their global searching abilities \cite{zeidler2001evolutionary},
\cite{zahedinejad2014evolutionary}, \cite{atabek2003evolutionary}. Among them, genetic algorithm (GA) has achieved great success in closed-loop learning control of laboratory quantum systems \cite{judson1992teaching}, \cite{tsubouchi2008rovibrational}. Differential evolution (DE) has emerged as another powerful technique in real optimization problems, especially
for robust control problems \cite{ma2015differential}, \cite{dong2020differential}. Evolutionary strategy (ES) methods have been applied in quantum control experiments \cite{shir2009niching} and exhibit advantage in exploring unknown environment in games \cite{salimans2017evolution}. Evolutionary algorithms are usually time consuming or space consuming when dealing with a group of population. In this paper, we will compare the performance of employing GD, DE, GA, ES to optimize a quantum autoencoder and their algorithm descriptions are outlined in \ref{app:pseudo code}.



\begin{rem}
The termination criterion for training the quantum autoencoder can be set as a maximum number of iterations, or the gap of $J(\textbf{u})$ between two successive iterations below a predefined small threshold.
\end{rem}



\section{Numerical results}\label{sec:numerical}
In this section, numerical results on 2-qubit and 3-qubit systems are presented. The system models and parameters are provided in Sec. \ref{subsec:parameters}. The input states are generated randomly and independently for several times to form a set of pure states with equal probabilities. In particular, a complex vector $|\psi_0\rangle=[a_1,a_2,....]^T$, $\sum_k|a_k|^2=1$  can be generated as $ [a_1,a_2,...]^T=[b_1+c_1\rm{i},b_2+c_2\rm{i},...]^T /\sqrt{\sum_{k=1} |b_k|^2+|c_k|^2}$ with the real vector $[b_1,b_2,...]^T$ and the  imaginary vector $[c_1,c_2,...]^T$ randomly sampled. Such input states are almost always linearly independent in the numerical simulation. For each compression task, numerical simulation is implemented for 20 runs independently, and each run deals with different input states. An example of compressing two 2-qubit states into two 1-qubit states is presented in Appendix G, with detailed information about the input states, encoded states and recovered states.
In subsection \ref{subsec:compression limit}, the performance learnt by
the closed-loop learning control approach is demonstrated to testify the upper
bound of the compression rate for given input states. Then, the performance of
four learning algorithms are compared and summarized in subsection
\ref{subsec:comparison}. In subsection \ref{subsec:influence}, the training
performance of the quantum autoencoder under different constraints is presented.


\subsection{System models and parameters}\label{subsec:parameters}
For a 2-qubit system, we assume that its Hamiltonian is described as $
H(t)=H_0+\sum_{j=1}^{4}u_j(t)H_j,$
where the free Hamiltonian is given as $H_0=\sigma_{z}\otimes \sigma_{z}$.
The control Hamiltonian operators are $H_1=\sigma_{x}\otimes I_2$, $H_2= I_2 \otimes \sigma_{x}$, $H_3=\sigma_{y}\otimes
I_2$, $H_4= I_2 \otimes \sigma_{y}$, respectively.
The control time duration $T_f=1.1$ a.u. is equally divided into $20$ sub-intervals. The
control fields are constrained as $u_1,u_2,u_3,u_4\in[-4,4]$.

For 3-qubit system, we assume that its Hamiltonian is $H(t)=H_0+\sum_{j=1}^{6}u_j(t)H_j$. Denote $\sigma_x^{(12)}=\sigma_x\otimes
\sigma_x \otimes I $, $\sigma_x^{(23)}= I\otimes \sigma_x \otimes \sigma_x$,
$\sigma_x^{(13)}= \sigma_x \otimes I \otimes \sigma_x $.  The free Hamiltonian is $
H_0=0.1\sigma_x^{(12)}+0.1\sigma_x^{(23)}+0.1\sigma_x^{(13)}$.
Denote $\sigma_x^{(1)}=\sigma_x\otimes I \otimes I $, $\sigma_x^{(2)}= I
\otimes \sigma_x \otimes I$, $\sigma_x^{(3)}= I \otimes I \otimes \sigma_x$,
and $\sigma_z^{(1)}=\sigma_z\otimes I \otimes I $, $\sigma_z^{(2)}= I \otimes
\sigma_z \otimes I$, $\sigma_z^{(3)}= I \otimes I \otimes \sigma_z$. The control Hamiltonian is $
       H_c(t)=u^{x}_{1}\sigma_x^{(1)}+u^{z}_{1}\sigma_z^{(1)}+u^{x}_{2}\sigma_x^{(2)}+u^{z}_2\sigma_z^{(2)}+u^{x}_3\sigma_x^{(3)}+u^{z}_{3}\sigma_z^{(3)}.$
The control time duration $T_f=20$ a.u. is equally divided into $100$ sub-intervals. The
control field is constrained as $u^{x}_{1}$, $u^{z}_{1}$, $u^{x}_{2}$, $u^{z}_{2}$,
$u^{x}_{3}$, $u^{z}_{3}$ $\in[0,1]$.

\subsection{Results on the maximum compression rate}\label{subsec:compression limit}
\begin{figure*}
	\centering
	\includegraphics[width=0.88\textwidth]{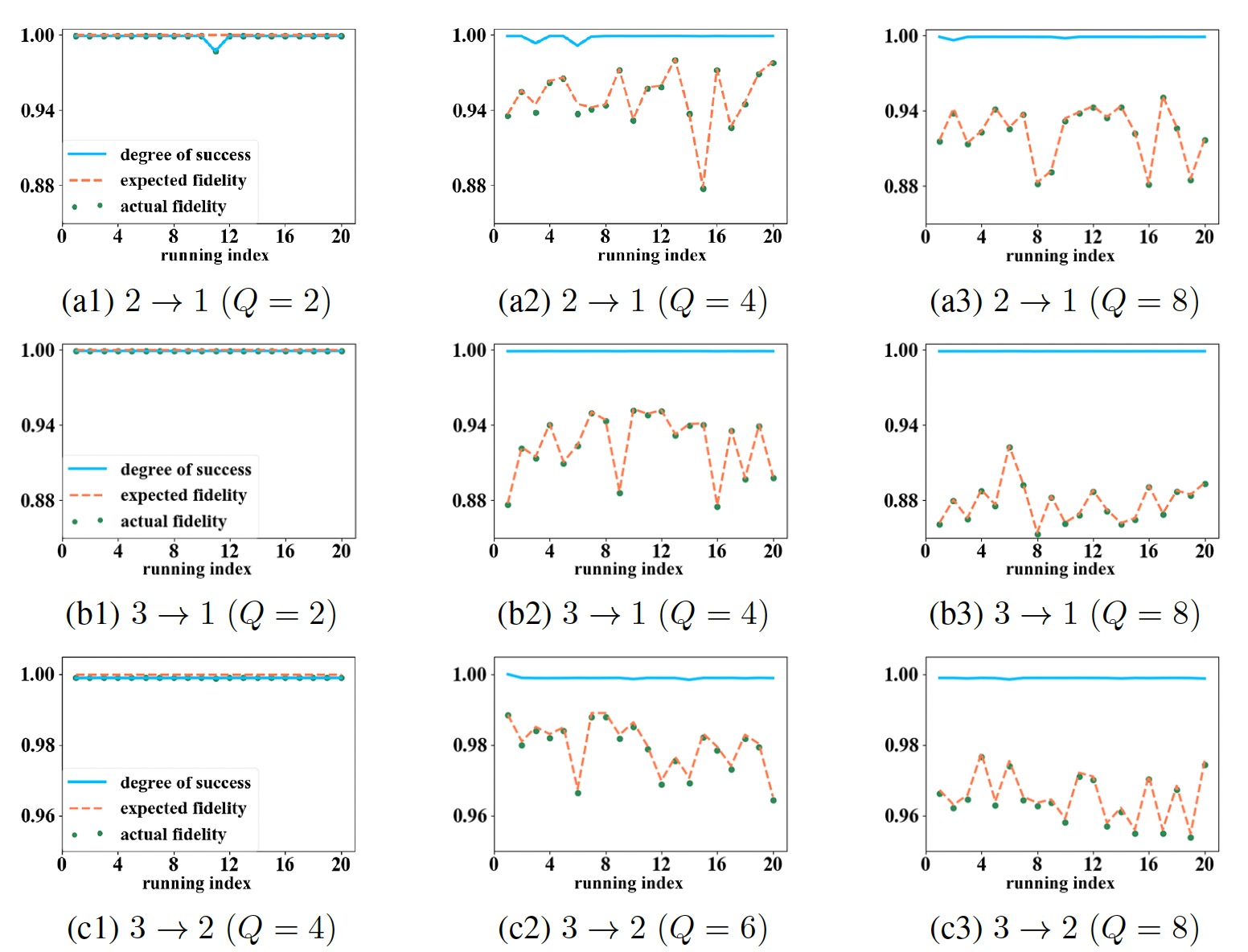}
	\caption{Relationship between the actual (best) fidelity found by learning algorithms and
		the expected optimal fidelity determined by the sum of partial eigenvalues.}
	\label{fig:relationship}
\end{figure*}

We present numerical results to demonstrate the effectiveness of the proposed learning method.
In Fig. \ref{fig:relationship}, the \emph{expected fidelity} is given
by the maximum sum of partial eigenvalues of the density operators based on (\ref{eq:eigens}),
while the \emph{actual fidelity} is given by the best training performance yielded by the learning methods. In addition, the \emph{degree of success} is defined as the ratio between the \emph{actual fidelity} and the \emph{expected fidelity}; i.e.,
$\emph{acutal fidelity} / \emph{expected fidelity}$. For simplicity, we denote ``$m\rightarrow n$ $(Q=k)$" as the task of compressing $m$-qubit states into $n$-qubit states, where the number of input states is $k$.

The results of 2-qubit input states (using GD) and 3-qubit input states (using ES) are presented in Fig. \ref{fig:relationship}, where the simulation for each compression task ``$m\rightarrow n$ $(Q=k)$" is implemented independently for 20 runs and different input states are used in each run. The degrees of success in most cases are almost $1$, revealing that the proposed approach is capable of reallocating the quantum information in the latent space with the optimal fidelities that they can theoretically achieve. A closer look at the subfigures of the first column reveals that, the learning method successfully finds the optimal control fields to preserve the quantum information within a lower dimensional space with nearly perfect compression rate for the tasks including $2\rightarrow 1$ $(Q=2)$, $3\rightarrow 1$ $(Q=2)$, and $3\rightarrow 2$ $(Q=4)$. For other cases, the actual fidelities are approaching the corresponding expected ones, resulting in approximate $100\%$ degree of success. The numerical result is consistent with the theoretical results in Corollary \ref{cor:limit value}.

\subsection{Performance comparison using different learning algorithms}\label{subsec:comparison}
\begin{figure*}[htb]
\centering
	\includegraphics[width=0.88\textwidth]{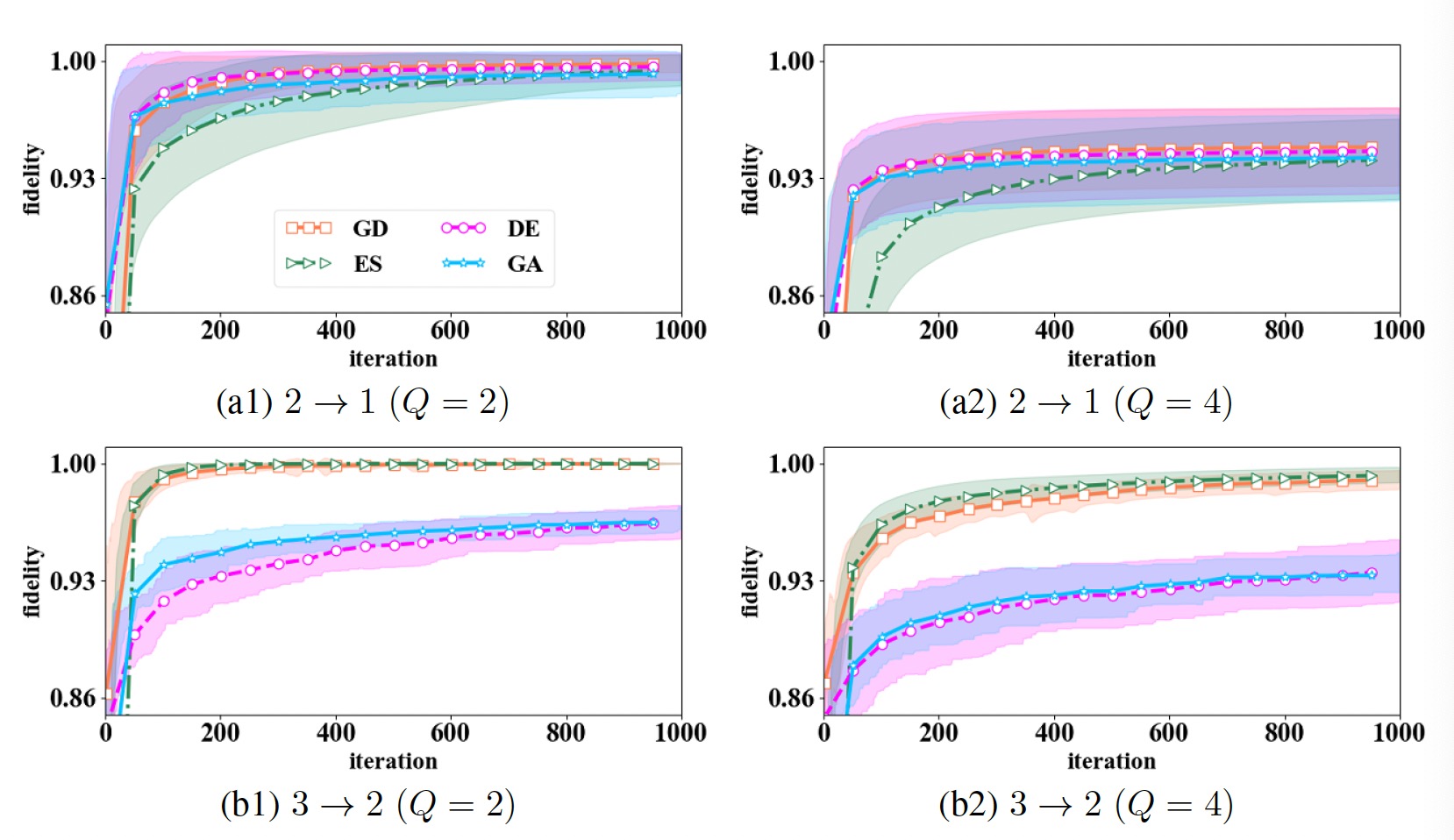}
\caption{Learning performance by four learning algorithms.}
\label{fig:four methods}
\end{figure*}

Here, we compare the performance of four learning algorithms on 2-qubit and 3-qubit systems.
Firstly, the training performance for compressing 2-qubit into 1-qubit and 3-qubit into 2-qubit states for 20 runs are shown in Fig. \ref{fig:four methods}. According to the results in Fig. \ref{fig:four methods} (a1) and (a2), the four algorithms achieve similar performance with almost the same fidelities and convergence rates on 2-qubit systems.
To be specific, DE is the fastest, while ES is the slowest. However, the performance of four algorithms on 3-qubit systems is quite different, where DE and GD exhibit comparative performance, while GA and DE fall far behind in Fig. \ref{fig:four methods} (b1) and (b2).

In addition, we measure the average recovered fidelities based on  (\ref{eq:J_recover}) using the optimal control strategies searched by the four learning algorithms. From Table \ref{tab:level4qubits2} and Table \ref{tab:level8qubits2}, the performance of GD regarding both mean and variance ranks first, with nearly perfect value for 2-qubit input states ($Q=2$) and 3-qubit input states ($Q=2,4$), closely followed by the results of ES. The high variances of DE and GA for 3-qubit systems reveal that they are not as robust as GD and ES, which achieve small variance values. The results show that GD and ES are more powerful and robust in the solving the optimization of the quantum autoencoder.

\begin{table}[htb]
\centering \caption{Statistical results of best fidelities for 2-qubit
systems}\label{tab:level4qubits2}
\scalebox{0.8}{
\begin{tabular}{c|cc|cc}
           & \multicolumn{2}{c}{$Q=2$} &  \multicolumn{2}{|c}{$Q=4$} \\
\hline
 Method &      Mean    &    Std    & Mean  & Std \\
 \hline
     GD    &  0.999275    &  3.142E-02  & 0.931266   & 2.696E-02   \\
 \hline
     ES    &  0.997923   &  8.444E-03  & 0.930307   & 2.755E-02   \\
 \hline
     DE    &  0.998163   &  7.544E-03  & 0.929607  & 2.847E-02  \\
 \hline
     GA    &  0.995779   &  9.492E-03  & 0.928438   & 2.974E-02
\end{tabular}
}
\end{table}

\begin{table}[htb]
\centering \caption{Statistical results of best fidelities for 3-qubit
systems}\label{tab:level8qubits2}
\scalebox{0.8}{
\begin{tabular}{c|cc|cc}
           & \multicolumn{2}{c}{$Q=2$} &  \multicolumn{2}{|c}{$Q=4$} \\
\hline
 Method &      Mean        &    Std       &   Mean      & Std \\
 \hline
     GD    &  0.999999  &  1.641E-12 & 0.999934  & 1.640E-04 \\
 \hline
     ES    &  0.999999  &  2.881E-10 & 0.999955  & 4.716E-05   \\
 \hline
     DE    &  0.993053  &  5.150E-03 & 0.948686  & 1.914E-02  \\
 \hline
     GA    &  0.974116  &  7.659E-03 & 0.927634  & 1.232E-02
\end{tabular}
}
\end{table}

\subsection{Variation of compression rate under different factors}\label{subsec:influence}


\begin{figure*}[htb]
	\centering
	\includegraphics[width=0.88\textwidth]{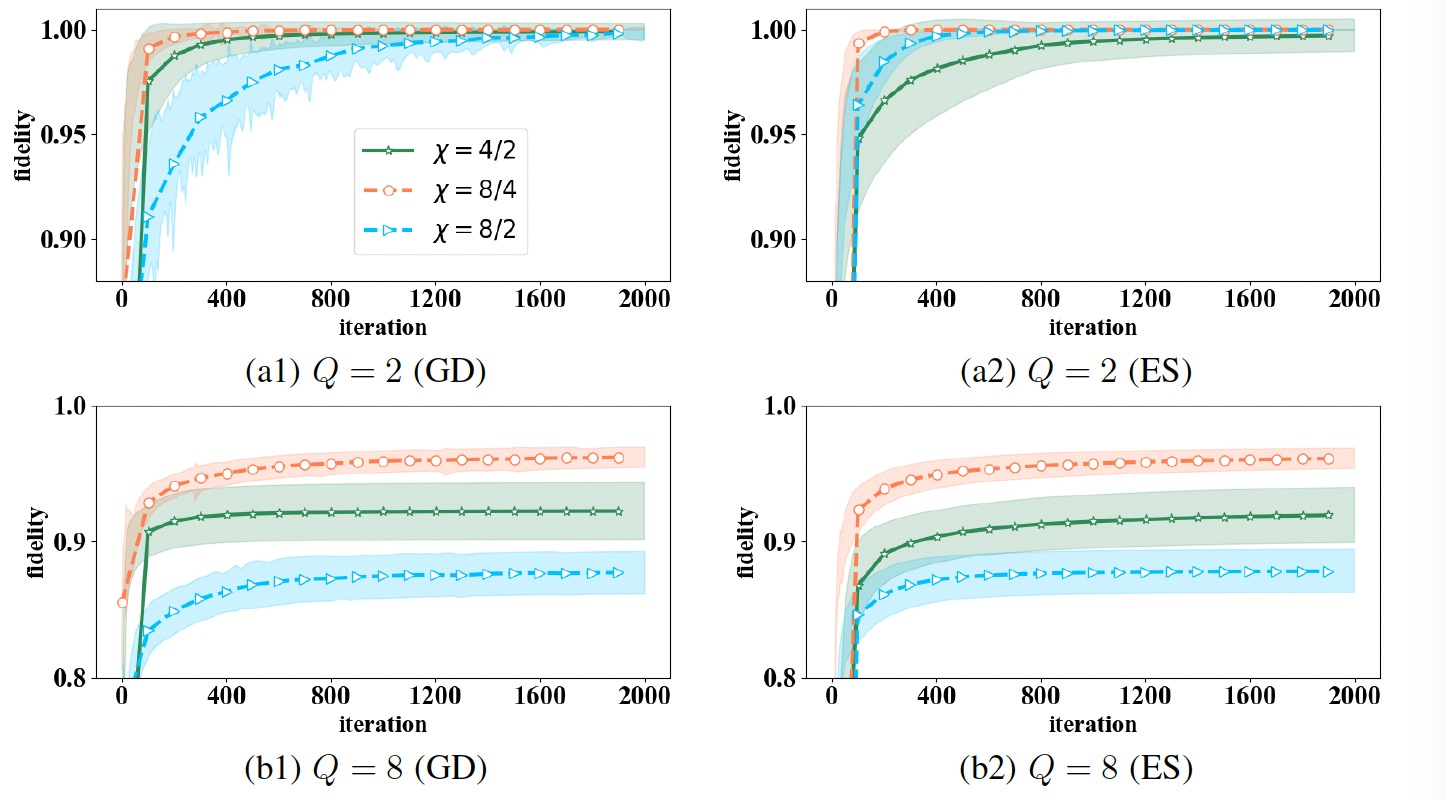}\caption{Learning performance for different number of input states via GD and ES.}
	\label{fig:increase dimension and trash}
\end{figure*}

Recall that $N$ and $N_B$ are the dimensions of the original input space and the latent space. We define $\chi=\frac{N}{N_B}$ as the compression ratio. A high value of $\chi$ means a tough compression task.
Based on the numerical comparison, ES method ranks first among three evolutionary algorithms and achieves excellent performance for different cases, especially for high dimensional cases.
Hence, we include the learning performance by GD and ES to explore their difference regarding different factors.

Firstly, we consider different input/latent dimensions.
In particular, $Q=2$ is illustrated as an easy case, since the theoretical upper
bound of the compression rate is $1$, with $Q=2\leq N_B$ for all cases. We also
include a relatively tough case, where the number of linearly independent vectors is
greater than the dimension of the latent space, considering $Q=8\geq N_B$ for
2-qubit and 3-qubit input states.
The learning performance of the two algorithms is demonstrated in Fig. \ref{fig:increase dimension and trash}. As we can see, both GD and ES achieve perfect fidelity $1$ with about 1000 iterations for the simple task. They also exhibit similar performance when 8 input states are simultaneously fed into a quantum autoencoder.
It is intuitive that the difficulty of training a quantum autoencoder increases with the compression ratio $\chi$, which is verified by the training results of GD and ES (see Fig. \ref{fig:increase dimension and trash} (a1), (b1) and (b2)), except that in Fig. \ref{fig:increase dimension and trash} (a2), where the learning performance of $\chi=\frac{4}{2}$ climbs with a slower speed compared with that of $\chi=\frac{8}{2}$.

\begin{figure}[htb]
\centering
\includegraphics[width=0.48\textwidth]{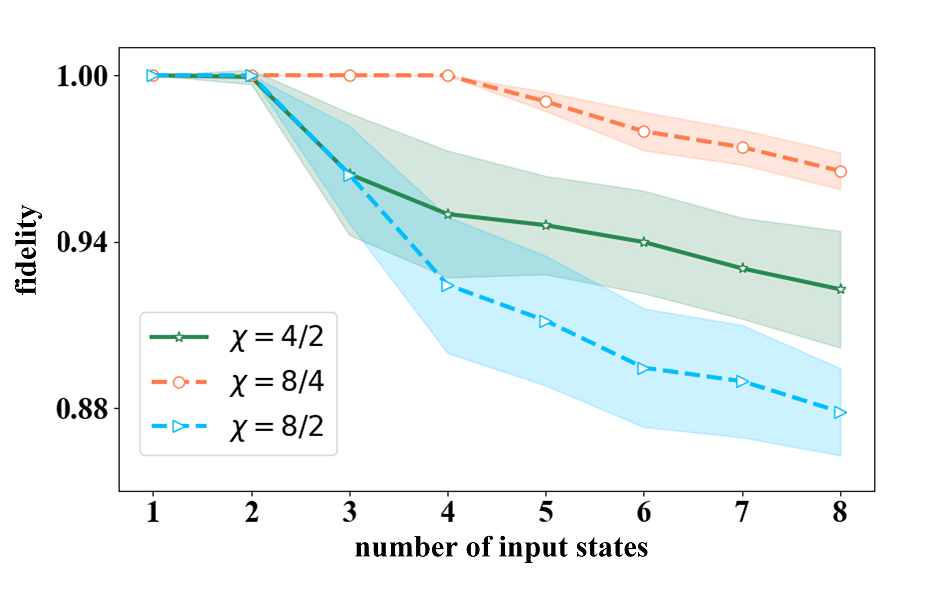}
\caption{Training performance of different input states under different $\chi$.}
\label{fig:increasing qubitsets}
\end{figure}

We also investigate the variation of the compression rate under different number of input states.
From Fig. \ref{fig:increasing qubitsets}, the training fidelity of $\chi=\frac{4}{2}$ remains around $1$ for $Q\leq2$, drops with increasing $Q$ and achieves an average value of $0.922810$ for $Q=8$. The numerical results of 3-qubit input states coincide with the theoretical demonstration, as perfect fidelity can be achieved when $Q\leq 4$ for compressing 3-qubit states into 2-qubit states or $Q\leq 2$ for compressing 3-qubit states into 1-qubit states. In particular, the compression task of 3-qubit states into 2-qubit states ($\chi=\frac{8}{4}$) exhibits a better  performance, as the values keep on the top and achieve an average value of $0.965500$ for $Q=8$. The task of compressing 3-qubit states into 1-qubit states ($\chi=\frac{8}{2}$) is demonstrated to be the most difficult, as the values drop sharply, ending up in an average of $0.878634$ at $Q=8$.

\section{Experimental results}\label{sec:experiments}

It is well known that any binary quantum alternative of a photon can serve
as a qubit \cite{wang2019quantum}. Thus, the polarization and path degrees of freedom can serve as two qubits.
In this section, we focus on the experimental realization of a quantum autoencoder to compress 2-qubit states into 1-qubit states.
Let $|R\rangle/|L\rangle$ be two eigenstates for path qubit and $|H\rangle/|V\rangle$ be two eigenstates for polarization qubit. Photons can be physical carriers of 2-qubit states, such as $|RH\rangle$ and $|LV\rangle$. Besides, 2-qubit universal parameterized unitary gates can be realized by combining path unitary gate with polarization gate \cite{englert2001universal}. Please refer to \ref{app:unitary gate} for detailed information.

\begin{figure}
\centering
\includegraphics[width=0.47\textwidth]{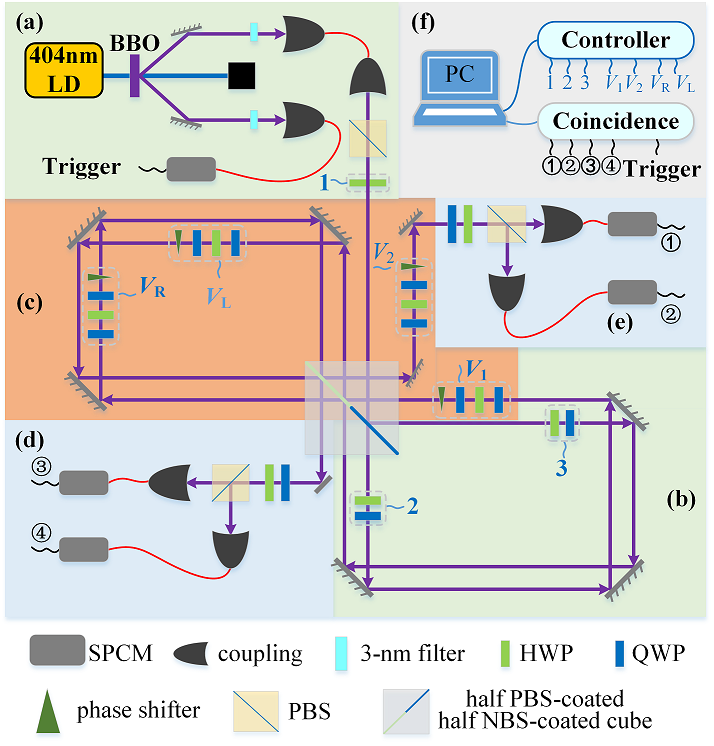}
\caption{The schematic of experimental setup for a quantum
autoencoder. The setup is divided into four modules.
(a)-(b) State preparation,
(c) Parameterized unitary gate,
(d)-(e) Measurements,
(f) Optimizing routine.}
\label{fig:device}
\end{figure}

The experimental setup is illustrated in Fig. \ref{fig:device}. It is divided into four modules. 1) \textbf{State preparation} (Fig. \ref{fig:device}. (a)-(b)): A Sagnac interferometer is used to generate phase-stable 2-qubit states. Photon pairs are created using type-I spontaneous parametric down-conversion
(SPDC) in a nonlinear crystal (BBO). One is sent to
 a single photon counting module (SPCM) to act as a trigger. The other is prepared in the state of highly
pure horizonal polarization state $|H\rangle$ through a
polarizer beam splitter (PBS). Then a half-wave plate
(HWP) along with a PBS controls the path qubit of
the photon. In each path, an HWP and a quarter-wave
plate (QWP) are used to control the polarization of the
photon. 2) \textbf{Parameterized unitary gate} (Fig. \ref{fig:device}. (c)):
Another interferometer is used to generate a 2-qubit unitary gate. In particular, four unitary polarization operators $V_1$, $V_2$, $V_R$ and $V_L$ are used, and each of them is composed of two QWPs,
an HWP, and a phase shifter (PS) consisting of a pair of wedge-shaped plates, which are controlled by a computer.
A special beam splitter cube that is half PBS-coated and half coated by a non-polarizer
beam splitter (NBS) is applied in the junction of two Sagnac
interferometers. 3) \textbf{Measurements} (Fig. \ref{fig:device}. (d)-(e)):
Local measurements on polarization can be achieved with the combination of a QWP, an HWP and a PBS. The typical count rate is set as 3000 photons per second. 4) \textbf{Optimization routine} (Fig. \ref{fig:device}. (f)):
A computer collects the coincidence and employs machine learning algorithms to optimize the rotations of the wave plates of the 2-qubit unitary gate.


The task is to find a 2-qubit unitary operator $U$ which can encode two 2-qubit states $|\psi_1\rangle, |\psi_2\rangle$ into two 1-qubit states $|\varphi_1\rangle,|\varphi_2\rangle$.
For example, $|RH\rangle$ and $|LV\rangle$ can be encoded as $|R\rangle |\varphi_1\rangle
$ and $|R\rangle|\varphi_2\rangle$.
In that case, path qubits can be discarded, with the original quantum information stored in the polarization qubits.
Similarly, encoding the information into path qubits is also feasible. We implement the experiments for compressing two 2-qubit states into two 1-qubit polarization states for two runs with different original states.
The input states of case 1 are $\frac{1}{\sqrt{2}}|RH\rangle+\frac{\rm{i}}{\sqrt{2}}|RV\rangle$ and $\frac{1}{\sqrt{2}}|LH\rangle+\frac{\rm{i}}{\sqrt{2}}|LV\rangle$.
The input states of case 2 are $\frac{1}{\sqrt{2}}|RH\rangle+\frac{\rm{i}}{\sqrt{2}}|LV\rangle$ and $\frac{1}{\sqrt{2}}|RV\rangle+\frac{\rm{i}}{\sqrt{2}}|LH\rangle$. The reference state is set as $|R\rangle$. The compression rate can be obtained by measuring the path qubits.

\begin{figure}
\centering
\includegraphics[width=0.48\textwidth]{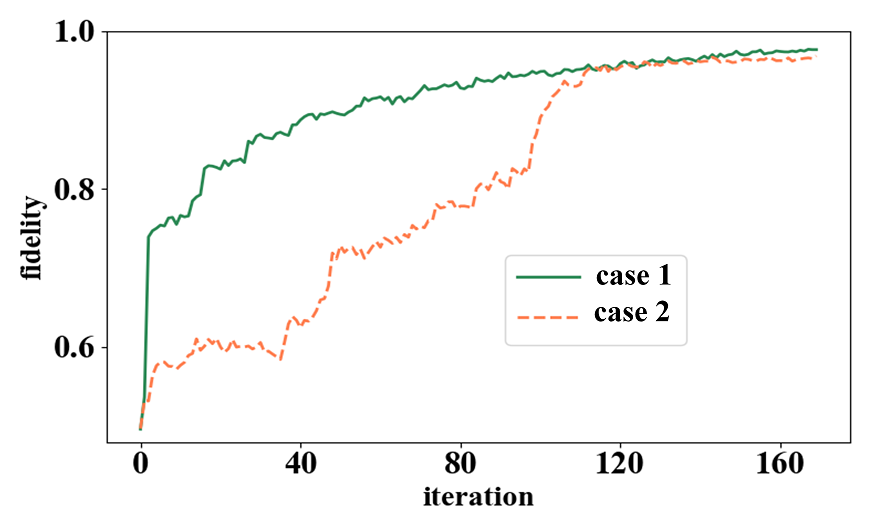}
\caption{The experimental results of encoding two 2-qubit states into two
1-qubit polarization states. Two lines correspond to two cases with different input states.}
\label{fig:experimental result}
\end{figure}

In particular, we apply a gradient algorithm in the closed-loop learning control approach to optimize the parameters of the device. The fidelities for two runs both reach near 1 within 160 iterations in Fig. \ref{fig:experimental result}, demonstrating the effectiveness of the proposed approach.

\section{Conclusion}\label{sec:conclusion}
We investigated the maximal compression rate of a quantum autoencoder and found that it is bounded by the maximum sum of partial eigenvalues of the density operator representation of the input states.
A closed-loop learning control framework was proposed to find the control fields for achieving the optimal compression rate, and numerical results were presented to show consistency results. The quantum optical experiment through a 2-qubit unitary gate to
compress two 2-qubit states into two 1-qubit states demonstrated the effectiveness of the quantum autoencoder.

Our future work will focus on the protocol of quantum autoencoders for mixed states and general quantum channels to deal with decoherence as well as experimental implementation on other platforms such as trapped ions and superconducting circuits. Considering that the required resource for measurement scales exponentially with the qubit number, it is meaningful to investigate the application of quantum autoencoders to reduce burden for quantum estimation and identification.

%

\textbf{Acknowledgements}
\\
\\
H. Ma, Y.  Wang, and D. Dong would like to thank the helpful discussions with Akira Sone.

\appendix
\renewcommand\thesection{\appendixname~\Alph{section}}
\renewcommand\theequation{\Alph{section}.\arabic{equation}}

\section{Proof for Lemma \ref{pro:disentanglement}}\label{app:disentanglement}
\textbf{Sufficiency}:
Given $U |\psi_{0}\rangle = |\psi_{A}\rangle \otimes |\psi_B\rangle=|\Psi_{\textup{ref}}\rangle \otimes |\psi_B\rangle$, it is clear that $J_1=F(U^{\dagger} \rho_{\textup{ref}} \otimes \textup{Tr}_{A}( U \rho_0 U^{\dagger}) U,|\psi_0\rangle \langle \psi_0|)=1$ and $J_2=F(\rho_{\textup{ref}}, \textup{Tr}_{B}(U \rho_0 U^{\dagger}) )=\langle \Psi_{\textup{ref}}| \Psi_{\textup{ref}}\rangle \langle \Psi_{\textup{ref}} |\Psi_{\textup{ref}}\rangle=1$.
\\
\textbf{Necessity}: 
Given $J_2=F(\rho_{\textup{ref}},\rho_A)=1$, we have $\rho_{\textup{ref}}=\rho_A$. Given $J_1=F(U^{\dagger} [\rho_{\textup{ref}} \otimes \textup{Tr}_{A}( U \rho_0 U^{\dagger})] U,\rho_0)$, we have $F(\rho_A \otimes \textup{Tr}_A(U\rho_0 U^{\dagger}), U \rho_0 U^{\dagger})=1$. Let $\rho_B=\textup{Tr}_A(U\rho_0 U^{\dagger})$, we obtain the following equation
\begin{equation}
\rho_A \otimes \rho_B=U\rho_0 U^{\dagger}.
\label{eq:appendix1}
\end{equation}
Since $\rho_0$ is a pure state, $\rho_A$  and $\rho_B$ must be pure states to guarantee that the eigenvalues of the left hand of (\ref{eq:appendix1}) are $1$ and $0$ with $(N-1)$ degeneracy. Ignoring the global phase, we have $U |\psi_0\rangle=|\psi_A\rangle  \otimes |\psi_B\rangle$ with $|\psi_A\rangle=|\Psi_{\textup{ref}}\rangle $ and $|\psi_B\rangle \langle \psi_B|=\textup{Tr}_A(U \rho_0 U^{\dagger})$. 



\section{Proof for Lemma \ref{pro:maximum fidelity}} \label{app:maximum fidelity}
The proof for Lemma \ref{pro:maximum fidelity} is drawn from \cite{jacobs2014quantum}, and is reorganized using the notation in this work.
\begin{pf}
	For a density operator $\rho$, there exists a unitary $W$ to diagonalize $\rho$ as
	$ W^{\dagger} \rho W=\textup{Diag}(\lambda_1,\lambda_2,\ldots,\lambda_N)$, where $\{\lambda_i\}$ are the eigenvalues of $\rho$ satisfying $\lambda_i
	\geq 0$ and $\sum
	\lambda_i =1$. For convenience, these eigenvalues are arranged in a descending order $\lambda_1\geq\lambda_2 \ldots \geq \lambda_{N} \geq 0$.
	Denote $D=\textup{Diag}(\lambda_1,\lambda_2,\ldots,\lambda_N)$, and we have the following relation:
\begin{equation}
\begin{split}
F(|\psi\rangle,\rho^U)=&\langle \psi| U \rho U^{\dagger} |\psi\rangle =\langle\psi| UW D W^{\dagger}U^{\dagger} |\psi\rangle \\
= &\langle \psi^{\prime}| D |\psi^{\prime}\rangle,
\label{eq:fidelities}
\end{split}
\end{equation}
	where $|\psi^{\prime}\rangle=W^{\dagger}U^{\dagger}|\psi\rangle$ is a pure
	state. Denote $|\psi^{\prime}\rangle$ in a vector representation as $|\psi^{\prime}\rangle=\left[b_1,b_2,\ldots,b_N\right]^T$ with $\sum_{i=1}^{N}|b_i|^2=1$.
	Then, Eq. (\ref{eq:fidelities}) can be rewritten as
	\begin{equation}
	F(|\psi\rangle,\rho^{U})=\left[b_1^{*},\ldots,b_N^{*}\right]
	\textup{Diag}(\lambda_1,...,\lambda_N)\left[b_1,\ldots,b_N\right]^{T}.
	\end{equation}
	Since $|b_i|^2 \geq 0$, we have $F(|\psi\rangle,\rho^{U})=\sum_i|b_i|^2 \lambda_i \leq \sum_i|b_i|^2 \lambda_1 =
	\lambda_1$.
	The optimal fidelity is achieved when
	$|\psi^{\prime}\rangle$ has the vector form as
	$|\psi^{\prime}\rangle=|e_{\lambda_{\max}}\rangle=[1,0,...,0]^{T}$. To
	achieve this, the unitary operator can be $U=\hat{U}^{\dagger}W^{\dagger}$, with $\hat{U}|\psi\rangle = |\psi^{\prime}\rangle$. Hence,
	$W$ can be obtained from the eigenvectors of $\rho$. The solution of $\hat{U}$ is given in \ref{app:unitary solution}.
	\hfill $\Box$
\end{pf}

\section{Diagonal elements of a reduced state}\label{app:partialtrace}
Let $\{|a_i\rangle\}$ be an orthogonal basis of $\mathbb{H}_A$ and $\{|b_i\rangle\}$ be an orthogonal basis of $\mathbb{H}_B$. A density operator $\rho_{AB}$ on $\mathbb{H}_A\otimes \mathbb{H}_B$ can be decomposed as $\rho_{A B}=\sum_{i j k l} c_{i j k l}|a_{i}\rangle\langle a_{j}|\otimes| b_{k}\rangle\langle b_{l}|$ and the partial trace reads: $
\operatorname{Tr}_{B} (\rho_{A B})=
\sum_{i j k l} c_{i j k l}|a_{i}\rangle \langle a_{j}|\textup{Tr} | b_{l} \rangle \langle b_{k}|
\\= \sum_{i j k l} c_{i j k l}|a_{i}\rangle \langle a_{j}|\langle b_{l} | b_{k}\rangle.$
Since $\{|b_i\rangle\}$ is a basis, we have $\rho_A=\textup{Tr}_{B}(\rho_{A B})=\sum_{i j k} c_{i j k k}|a_{i}\rangle \langle a_{j}|$. Notice that the diagonal elements of $\rho_A$ correspond to the coefficients of $|a_i\rangle \langle a_j|$ with $i=j$.
It is clear that $[\rho_A]_{pp}=\sum_{k} c_{p p k k}.$
Since $\rho_{AB}$ can be written as $\sum_{i j k l} c_{i j k l}|a_{i}\rangle | b_{k}\rangle \otimes \langle a_{j}|\langle b_{l}|$.
The coefficients $c_{ijkl}$ for $i=j=p$ and $l=k$ contribute to the diagonal elements of $\rho_{AB}$. For $p=1$, the first diagonal element of $\rho_{AB}$ is given as $\sum_{k} c_{1 1 k k}.$

\section{Solution of unitary transformation}\label{app:unitary solution}
Given two vectors $|x\rangle \in \mathbb{C}^{N}$ and $|y\rangle \in \mathbb{C}^{N}$, the goal is to
find a unitary $U$ that satisfies $U|x\rangle=|y\rangle$.
Here, we adopt the complex \emph{Householder} transformation, which can realize a state transition from a complex vector $|x\rangle$  to another complex vector $|y\rangle$. The unitary solution for the state transition is drawn from \cite{venkaiah1993householder}, and is reorganized using the notation in this work.

Let $H=I_{N}-k|w\rangle \langle w|$, where
$|w\rangle=(|x\rangle-|y\rangle)/||x\rangle-|y\rangle
|$, and $k=\frac{2-\langle y
	|x\rangle-\langle x |y\rangle}{1-\langle y |x\rangle}$.
$H$ can be rewritten as
$H=I_N-\frac{(|x\rangle-|y\rangle) (\langle x| -\langle y|)}{1-\langle
	y|x\rangle}$. A simple calculation reveals that $H|x\rangle=|y\rangle$.
Additionally, we have $HH^{\dagger}=I_N+(|k|^{2}-k-k^*)|w\rangle
\langle w|$, and
\begin{equation}
\begin{split}
k+k^{*}=&\frac{2-\langle x |y\rangle-\langle y |x\rangle}{1-\langle x
	|y\rangle}+
\frac{2-\langle x |y\rangle-\langle y |x\rangle}{1-\langle y |x\rangle}\\
=&\frac{(2-\langle x |y\rangle-\langle y |x\rangle)(2-\langle x
	|y\rangle-\langle y |x\rangle)}{(1-\langle x |y\rangle)(1-\langle y
	|x\rangle)}\\
=&|k|^{2}.
\end{split}
\end{equation}
That is $HH^{\dagger}=I_N$.
Hence, we have $H|x\rangle=y$ and $HH^{\dagger}=I_{N}$.

As such, a unitary transformation $\bar{U}=I_N-\frac{(|x\rangle-|y\rangle) (\langle x| -\langle y|)}{1-\langle y|x\rangle}$ transforms $|x\rangle$ to $|y\rangle$.
Denote $|x\rangle=|z_0\rangle$ and $|y\rangle=|z_{M+1}\rangle$. Choosing different intermediate points $|z_{1}\rangle, |z_{2}\rangle,...,|z_{M}\rangle$, we adopt the same method to figure out an individual unitary for each state transition, i.e., $U_k|z_k\rangle=|z_{k+1}\rangle$ ($k=1,2,...,M$). Hence, we obtain a sequence, $|z_0\rangle \xrightarrow{U_0} |z_{1}\rangle...|z_{M}\rangle \xrightarrow{U_M}|z_{M+1}\rangle$, and the overall unitary transformation  $U=\prod_{k=0}^{M} U_{k}$ achieves $U|x\rangle=|y\rangle$.

\section{Pseudo code for machine learning algorithms}\label{app:pseudo code}
Here, we provide the pseudo code for machine learning algorithms that are used in the main text.
Denote the control fields as a column vector $\theta$ and let $J(\theta)$ be the loss function to be optimized.
The goal of the algorithms is to find the optimal vector $\theta^{*}$ such that $J(\theta^{*})$ achieves the optimal value.
Considering the physical restriction of control fields, the parameters are usually initialized as
\begin{equation}
\theta=\textbf{u}_{\textup{min}} +\textup{Rand}[0,1](\textbf{u}_{\textup{max}}-\textbf{u}_{\textup{min}}),
\label{eq:initialization}
\end{equation}
where $\textup{Rand}[0,1]$ is function generating random numbers uniformly distributed between 0 and 1. This method is usually adopted to randomly initialize parameters to guarantee enough diversity. Besides, boundary check and resetting values are required for every step that involves new vectors to guarantee that newly generated vectors lie in the constrained field.



Given the learning rate $\alpha$, the main procedure of GD is as follows:\\
Step 1. Randomly initialize the parameters based on (\ref{eq:initialization});\\
Step 2. Compute the gradient information $\frac{\partial J(\theta)}{\partial{\theta}}$;\\
Step 3. Update the parameter $\theta \leftarrow \theta + \alpha \frac{\partial J(\theta)}{\partial{\theta}}$.\\

Here, the gradients are approximated by finite difference gradient estimator \cite{pepper2019experimental}. That is to perform tiny perturbations on every direction of the vector.  In particular, we may generate a unit vector $v^j=[0,...,0,1,0,...,0]^T$, where only the $j-$th element is one. Then, two new vectors $\theta_{+}^j=\theta+\beta v^j$ and $\theta_{-}^j=\theta-\beta v^j$ are obtained to approximate the $j$-th element of gradients as $\frac{\partial J(\theta)}{\partial{\theta^j}}\cong\frac{J(\theta^j_{+})-J(\theta^j_{-})}{2\beta}$, where $\beta$ is a perturbation factor. 
Repeating the process with different choices of $j$ among $\{1,2,...\}$, the gradients read as $\frac{\partial J(\theta)}{\partial{\theta}}=[\frac{\partial J(\theta)}{\partial{\theta_1}},...,\frac{\partial J(\theta)}{\partial{\theta_j}},...]^T$.


Given the population size $NP$, the crossing-over rate $P_c$, the mutation rate $P_m$, the main procedure of GA is as follows.\\
Step 1. Randomly generate $NP$ individuals $\{\theta_i\}$ based on (\ref{eq:initialization}) and constitute $S=\{\theta_1,...,\theta_{NP}\}$;\\
Step 2. Rank $\{\theta_i\in S\}$ according to $\{J(\theta_i)\}$ (descending);\\
Step 3. Select top $\lceil  NP(1-P_c)\rceil$ vectors to constitute $S_{1}$;\\
Step 4. Sample $NP-\lceil  NP(1-P_c)\rceil$ vectors from $S$ with
probability $P(\theta_i)=\frac{J(\theta_i)}{\sum^{NP}_{j=1}J(\theta_j)}$, to constitute $S_{2}$;\\
Step 5. Randomly pair vectors among $S_{2}$ and perform $\lceil N_2/2 \rceil$ times of crossover to renew the vectors in $S_{2}$;\\
Step 6. Mutate vectors in $S_{2}$ with probability $P_m$;\\
Step 7. Obtain new generation $S \leftarrow \{S_{1},S_{2}\}$;\\
Step 8. If convergent, go to Step 9, otherwise go to Step 2;\\
Step 9. Optimal control parameters $\theta^*= {\arg\max}_{\theta_i}(J(\theta_i))$.

Given the population size $NP$, the scaling factor $F$ and the cross-over ratio $CR$, the main procedure of DE is as follows.\\
Step 1. Randomly generate $NP$ individuals $\{\theta_i\}$ based on (\ref{eq:initialization});\\
Step 2. Repeat for each individual $i=1,...,NP$\\
Step 2-1. Generate three distinct indices $(i_1,i_2,i_3)\neq i$;\\
Step 2-2. Perform mutation $u_i=\theta_{i_1}+F \cdot (\theta_{i_2} - \theta_{i_3})$;\\
Step 2-3. Perform cross-over on $u_i$ to obtain $v_i$ using binomial (uniform) crossover with $CR$;\\
Step 2-4. Perform selection $\theta_{i} \leftarrow \mathop{\arg\max}_{x\in \{\theta_i,v_i\}} J(x)$;\\
Step 3. If convergent, go to Step 4, otherwise go to Step 2;\\
Step 4. Optimal control parameters $\theta^*= \mathop{\arg\max}_{\theta_i}(J(\theta_i))$.\\

Given the population size $NP$, the learning rate $\alpha$, the momentum coefficient $\beta$, and the permutation factor $\delta$, the main procedure of ES is as follows.\\
Step 1. Initialize gradient $dJ=0$ and momentum $dv=0$;\\
Step 2. Initialize the mean vector $\bar{\theta}$ according to (\ref{eq:initialization});\\
Step 3. Repeat for each individual $i=1,...,NP$\\
Step 3-1. Sample variation $\epsilon_{i} \sim N(0,I)$;\\
Step 3-2. Set mutation variant as $X_i \leftarrow \bar{\theta}+ \delta \epsilon_i $;\\
Step 4. Obtain gradient $ dJ \leftarrow \frac{1}{NP\delta}\sum_{j=1}^{NP}J(X_i)\epsilon_j$;\\
Step 5. Obtain momentum $dv \leftarrow \beta dv +(1-\beta) dJ;$\\
Step 6. Update the new mean vector $\bar{\theta} \leftarrow \bar{\theta} + \alpha dv $;\\
Step 7. If convergent, go to Step 8; otherwise go to Step 3;\\
Step 8. Optimal control parameters $\theta^*=\bar{\theta}$.

\section{Universal two-qubit unitary gate}\label{app:unitary gate}
The setup for generating a universal two-qubit unitary gate is composed of a path unitary gate and a polarization gate \cite{englert2001universal}.
The 2-qubit unitary gate $U$ can be written as $U=\left[\begin{array}{cc}{U_{RR}} & {U_{RL}} \\ {U_{LR}} & {U_{LL}}\end{array}\right] $,
where $U_{RR},U_{RL},U_{LR}$ and $U_{LL}$ are $2\times 2$ matrices referring to the path $R/L$ alternative. They have the form of $U_{RR}=\frac{1}{2}V_2(V_R+V_L)V_1$, $U_{LL}=\frac{1}{2}(V_R+V_L)$, $U_{RL}=-\frac{\rm{i}}{2}V_2(V_R-V_L)$, and $U_{LR}=\frac{\rm{i}}{2}(V_R-V_L)V_1$.
Thus, one may find 4 unitary polarization operators $V_1,V_2,V_R$ and $V_L$ to achieve any 2-qubit unitary operator by using a set of QWPs, HWPs and phase shifters.

\section{A numerical example of quantum autoencoders}

Here, we provide an example of compressing two 2-qubit states into two 1-qubit states with $|\psi_{\textup{ref}}\rangle =[1,0]^T$. The input states are given as \\
	\begin{equation}
	\resizebox{.95\hsize}{!}{$
		|\psi_0^1\rangle=[0.3497+0.0030\rm{i},0.1791+0.0782\rm{i},0.2732+0.4317\rm{i},0.5438+0.5316\rm{i}]^T,
		$}
	\nonumber
	\end{equation}
	\begin{equation}
	\resizebox{.95\hsize}{!}{$
		|\psi_0^2\rangle=[0.0913+0.1238\rm{i},0.3842+0.0724\rm{i},0.5955+0.1468\rm{i},0.1398+0.6539\rm{i}]^T.
		$}
	\nonumber
	\end{equation}
	Their density matrix representations are given as
	\begin{equation}
	\resizebox{.95\hsize}{!}{$
		\rho_0^1=\left[\begin{array}{cccc}
		{0.1224} &{ 0.0901 - 0.0008\rm{i}} & {0.0969 + 0.1502\rm{i}} & {0.1918 + 0.1843\rm{i}}\\
		{0.0901 + 0.0008\rm{i}}  & {0.0663} & {0.0703 + 0.1112\rm{i}} &{0.1400 + 0.1369\rm{i}}\\
		{0.0969 - 0.1502\rm{i}} & {0.0703 - 0.1112\rm{i}} & {0.2611 } & {0.3781 - 0.0895\rm{i}}\\
		{0.1918 - 0.1843\rm{i}} & {0.1400 - 0.1369\rm{i}} & {0.3781 + 0.0895\rm{i}} & {0.5783}\\
		\end{array}\right],
		$}
	\nonumber
	\end{equation}
	\begin{equation}
	\resizebox{.95\hsize}{!}{$
		\rho_0^2=\left[\begin{array}{cccc}
		{0.0237} &{ 0.0441 - 0.0410\rm{i}} & {0.0726 - 0.0603\rm{i}} & {0.0937 + 0.0424\rm{i}}\\
		{0.0441 + 0.0410\rm{i}}  & {0.1529 } & {0.2395 + 0.0133\rm{i}} &{0.1011 + 0.2412\rm{i}}\\
		{0.0726 + 0.0603\rm{i}} & {0.2395 - 0.0133\rm{i}} & {0.3763} & {0.1793 + 0.3689\rm{i}}\\
		{0.0937 - 0.0424\rm{i}} & {0.1011 - 0.2412\rm{i}} & {0.1793 - 0.3689\rm{i}} & {0.4471}\\
		\end{array}\right].
		$}
	\nonumber
	\end{equation}
	By optimizing the quantum autoencoder using ES, the latent states are obtained as \\
	\begin{equation}
	\resizebox{.7\hsize}{!}{$
		\rho_B^1=\left[\begin{array}{cc}{0.2142} &{0.1577+0.3786\rm{i}} \\ {0.1577-0.3786\rm{i}}  & {0.7858}\end{array}\right],
		$}	
	\nonumber
	\end{equation}
	\begin{equation}
	\resizebox{.7\hsize}{!}{$
		\rho_B^2=\left[\begin{array}{cc}{0.0778} & {0.0937-0.2506\rm{i}} \\ {0.0937+0.2506\rm{i}} & {0.9222}\end{array}\right].
		\nonumber$}
	\end{equation}
	The recovered states are obtained as
	\begin{equation}
	\resizebox{.95\hsize}{!}{$
		\rho_f^1=\left[\begin{array}{cccc}
		{0.1283} &{0.0666-0.0263\rm{i}} & {0.0976-0.1546\rm{i}} & {0.1931-0.1896\rm{i}}\\
		{0.0666+0.0263\rm{i}}  & {0.0400} & {0.0825-0.0602\rm{i}} &{0.1392-0.0588\rm{i}}\\
		{0.0976+0.1546\rm{i}} & {0.0825+0.0602\rm{i}} & {0.2608} & {0.3755+0.0885\rm{i}}\\
		{0.1931+0.1896\rm{i}} & {0.1392+0.0588\rm{i}} & {0.3755-0.0885\rm{i}} & {0.5708}\\
		\end{array}\right],
		$}
	\nonumber
	\end{equation}
	\begin{equation}
	\resizebox{.95\hsize}{!}{$
		\rho_f^2=\left[\begin{array}{cccc}
		{0.0239} &{0.0417+0.0412\rm{i}} & {0.0712 +0.0637\rm{i}} & {0.0961-0.0391\rm{i}}\\
		{0.0417-0.0412\rm{i}}  & {0.1438} & {0.2341-0.0115\rm{i}} &{0.1002-0.2338\rm{i}}\\
		{0.0712-0.0637\rm{i}} & {0.2341+0.0115\rm{i}} & {0.3821} & {0.1819-0.3726\rm{i}}\\
		{0.0961+0.0391\rm{i}} & {0.1002+0.2338\rm{i}} & {0.1819+0.3726\rm{i}} & {0.4501}\\
		\end{array}\right].
		$}
	\nonumber
	\end{equation}

\bibliographystyle{alpha}

\end{document}